# Spectral Response of SPHEREx

Howard Hui[1,2], James J. Bock[1,2], Samuel Condon[3], C. Darren Dowell[2,1], Woong-Seob Jeong[4], Young-soo Jo[4], Phil M. Korngut[1,2], Kenneth Manatt[2], Chi Nguyen[1], Hien Nguyen[2,5], Stephen Padin[1], Sung-Joon Park[4], Jeonghyun Pyo[4], Yujin Yang[4], Matthew L. N. Ashby[6], Yoonsoo P. Bach[4], Tzu-Ching Chang[2,1], Yun-Ting Cheng[1,2], Yi-Kuan Chiang[7], Asantha Cooray[8], Brendan P. Crill[2,1], Ari J. Cukierman[1], Olivier Doré[2,1], Andreas L. Faisst[9], Joseph L. Hora[6], Jae Hwan Kang[1], BoMee Lee[4,9], Carey M. Lisse[10], Daniel C. Masters[9], Roberta Paladini[9], Zafar Rustamkulov[9], Volker Tolls[6], Michael W. Werner[2], and Michael Zemcov[11,2]

[1] Department of Physics, California Institute of Technology, 1200 East California Boulevard, Pasadena, CA 91125, USA; hhui@caltech.edu
[2] Jet Propulsion Laboratory, California Institute of Technology, 4800 Oak Grove Drive, Pasadena, CA 91109, USA
[3] Department of Physics, Stanford University, Stanford, CA 94305, USA
[4] Korea Astronomy and Space Science Institute (KASI), 776 Daedeok-daero, Yuseong-gu, Daejeon 34055, Republic of Korea
[5] University of Science, Viet Nam National University Ho Chi Minh City, 227 Nguyen Van Cu, Ho Chi Minh City, 700000, Vietnam
[6] Center for Astrophysics | Harvard & Smithsonian, Optical and Infrared Astronomy Division, Cambridge, MA 01238, USA
[7] Academia Sinica Institute of Astronomy and Astrophysics (ASIAA), Number 1, Section 4, Roosevelt Road, Taipei 10617, Taiwan
[8] Department of Physics & Astronomy, University of California Irvine, Irvine, CA 92697, USA
[9] IPAC, California Institute of Technology, MC 100-22, 1200 East California Boulevard, Pasadena, CA 91125, USA
[10] Johns Hopkins University Applied Physics Laboratory, Laurel, MD 20723, USA
[11] School of Physics and Astronomy, Rochester Institute of Technology, 1 Lomb Memorial Drive, Rochester, NY 14623, USA

## Abstract

The Spectro Photometer for the History of the Universe, Epoch of Reionization, and Ices Explorer (SPHEREx) is conducting the first all-sky near-infrared spectral survey spanning 0.75–5.0 $\mu$m with resolving power $R \approx 35$–130. Linear variable filters mounted in front of six H2RG detectors produce a position-dependent spectral response across the focal plane. This paper presents the ground-based spectral calibration of SPHEREx, including the cryogenic apparatus, optical configuration, measurement strategy, analysis pipeline, and resulting calibration products. Monochromatic wavelength scans are used to derive the spectral response function, band center, and resolving power for every pixel. Band centers are measured to better than 1 nm for Bands 1 through 4 (0.75–3.82 $\mu$m) and better than 10 nm for Bands 5 and 6 (3.82–5.0 $\mu$m). Out-of-band leakage is negligible for detectors above 1.64 $\mu$m and is present at the percent level below this wavelength. The resolving power is measured to within 5% and agrees with design expectations to within 10%. An on-sky spectrum of the Cat's Eye Nebula (NGC 6543) constructed from repeated observations provides in-flight verification and shows agreement between ground-calibrated response and astrophysical emission features. Calibration products, including per-pixel band center and resolving power maps, are released through IPAC to support community use of SPHEREx data. The absolute spectral calibration will continue to improve through in-flight measurements, with further reductions in uncertainty expected for the longest-wavelength bands.

## 1. Introduction

The Spectro Photometer for the History of the Universe, Epoch of Reionization, and Ices Explorer (SPHEREx) is a NASA Astrophysics Medium Class Explorer mission that launched on 2025 March 11 (J. J. Bock et al. 2026). SPHEREx is conducting the first all-sky near-infrared spectral survey, covering 0.75–5.0 $\mu$m with resolving power $R \approx 35$–130. The instrument produces spectra in 102 spectral channels for every 6.″15 pixel on the sky, enabling investigations in cosmology, galaxy evolution, and Galactic astrophysics (O. Dore et al. 2018). Key science goals include constraining inflationary physics through large-scale structure, probing the origin of the extragalactic background light, and mapping biogenic ices in the interstellar medium.

The science goals require precise preflight calibration of the instrument. Critical tests include detector and readout characterization (C. H. Nguyen et al. 2025), focus (S. S. Condon et al. 2024), and measurement of the full spectral response for each of the approximately 25 million pixels. A summary of mission-level integration and testing is provided in P. M. Korngut et al. (2024). Spectral calibration is particularly important because it cannot easily be done in orbit with the same spectral coverage or control. Accurate knowledge of the pixel-level wavelength response and resolving power is essential for determining redshifts, measuring absorption features, and modeling astrophysical foregrounds. Initial calibration results were presented in H. Hui et al. (2024). The present work describes the complete ground spectral calibration effort and the full spectral analysis pipeline, which produces a spectral data product that feeds into the main data reduction pipeline described in R. Akeson et al. (2025).

Section 2 provides an overview of the SPHEREx optical layout and spectrometer design, and describes the band definition and wavelength progression across each detector set by the linear variable filters (LVFs). Section 3 describes the calibration hardware, cryogenic test beds, and measurement sequence. Section 4 outlines the data reduction pipeline. Section 5 discusses measurement systematics and quantifies the calibration uncertainty. Section 6 reports the derived spectral parameters and final calibration products. Section 7 presents conclusions and implications for in-flight operations.

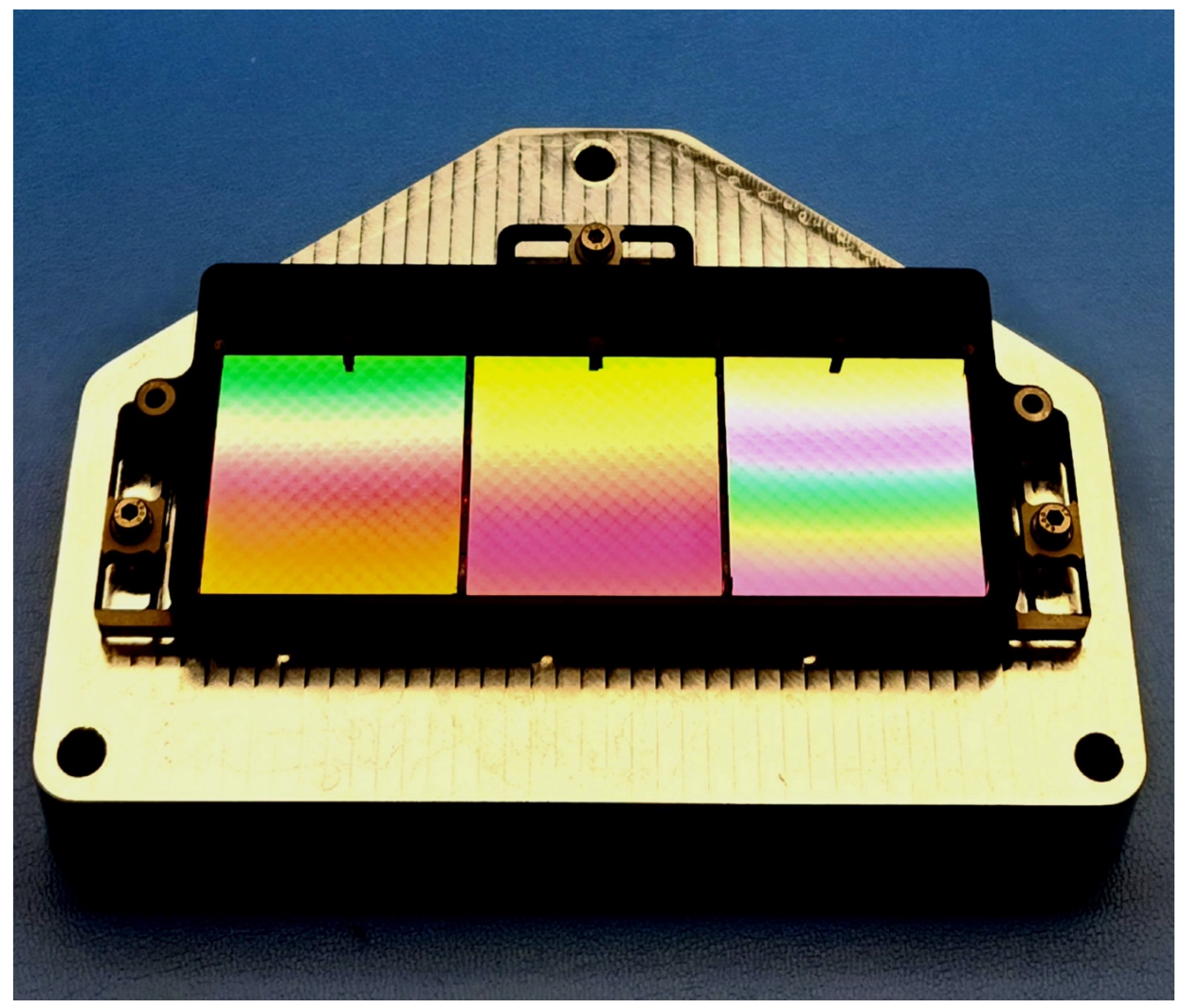

**Figure 1.** MWIR focal plane assembly containing three H2RG detectors, each covered by an LVF (P. M. Korngut et al. 2024). The apparent color gradient arises from interference effects at the LVF surface. The effective wavelength increases monotonically along the vertical direction in the image. The SPHEREx instrument includes two such focal plane assemblies, one for the SWIR channel and one for the MWIR channel.

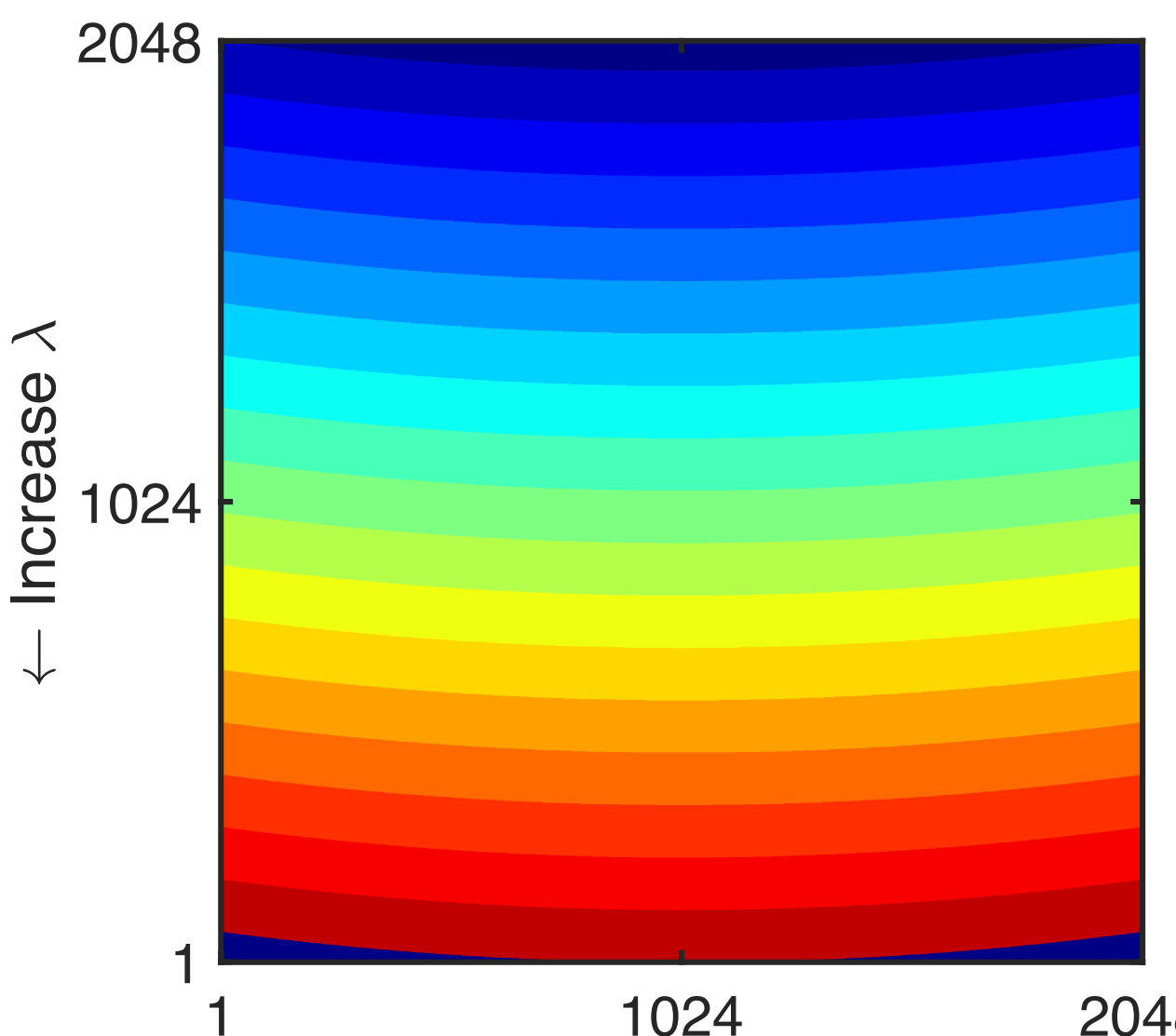

**Figure 2.** Spectral channel layout for each detector. Channels are defined geometrically rather than by direct wavelength measurements to simplify survey planning. The LVF produces a monotonic wavelength gradient along the vertical axis. Curvature near the band edges causes a subset of pixels to fall outside the nominal 17-channel layout, and these pixels provide cross-band calibration information.

## 2. SPHEREx Instrument Overview

The SPHEREx instrument uses a compact three-mirror anastigmat telescope with a 20 cm aperture and a $3\overset{\circ}{.}5 \times 11\overset{\circ}{.}3$ field of view (J. J. Bock et al. 2026). After reflection from the third mirror, a dichroic beam splitter divides the incoming light into two wavelength ranges. Wavelengths shorter than 2.42 $\mu$m are reflected to the short-wavelength infrared (SWIR) focal plane assembly, and longer wavelengths are transmitted to the midwavelength infrared (MWIR) focal plane assembly.

Each focal plane assembly contains three 2048 × 2048 H2RG detectors (R. Blank et al. 2011). The outermost four rows and columns are dark reference pixels, leaving 2040 × 2040 optically active pixels per array. An LVF is mounted directly above each detector (Figure 1), producing a smoothly varying, position-dependent spectral bandpass across the focal plane. The focal plane therefore functions as a spectral imager in which wavelength is encoded as a function of pixel position.

An LVF is a bandpass interference filter whose central wavelength varies along one spatial axis (K. P. Rosenberg et al. 1994). Each of the six detector arrays is paired with a distinct LVF, defining six spectral bands that together span the full SPHEREx wavelength range. For survey planning purposes, each band is divided into 17 nominal spectral channels, resulting in a total of 102 channels across the instrument (Figure 2). Over the course of the survey, each $6\overset{''}{.}15$ sky pixel is observed in each of these 102 spectral channels at least four times, and the resolving power of each channel, combined with the survey scan strategy, ensures approximate Nyquist sampling of the spectrum over the full sky by the end of the mission (S. Bryan et al. 2025).

The curvature of the spectral channels arises from the LVF fabrication process. This effect causes a subset of pixels to fall outside the nominal 17-channel layout. These pixels are fully optically active and exhibit a continuous wavelength progression consistent with the main channels, and their wavelength coverage overlaps with adjacent detector bands. Although they are not required for survey completeness, they provide valuable cross-band consistency checks. Table 1 lists the design wavelength coverage and the nominal resolving power for each band.

The channel boundaries for the $i$th channel in a given band are

$$\lambda_{\min,i} = \lambda_{\min}\left(\frac{\lambda_{\max}}{\lambda_{\min}}\right)^{\frac{i-1}{17}},\quad i = 1 \ldots 17, \tag{1}$$

$$\lambda_{\max,i} = \lambda_{\min}\left(\frac{\lambda_{\max}}{\lambda_{\min}}\right)^{\frac{i}{17}},\quad i = 1 \ldots 17, \tag{2}$$

where $\lambda_{\min}$ and $\lambda_{\max}$ are the design minimum and maximum band centers $\lambda_c$ as shown in Table 1.

For each pixel, the band center $\lambda_c$ is defined as the wavelength at which the cumulative integral of the spectral response $T_\lambda$ reaches 50%:

$$\int_0^{\lambda_c} T_\lambda\, d\lambda = \frac{1}{2}\int_0^{\infty} T_\lambda\, d\lambda. \tag{3}$$

This definition captures the centroid of the non-Gaussian bandpass at shorter wavelengths, and incorporates the contribution from out-of-band leakage into the band center, providing a more representative effective wavelength for each pixel.

The resolving power $R$ is

$$R = \frac{\lambda_c}{\Delta\lambda}, \tag{4}$$

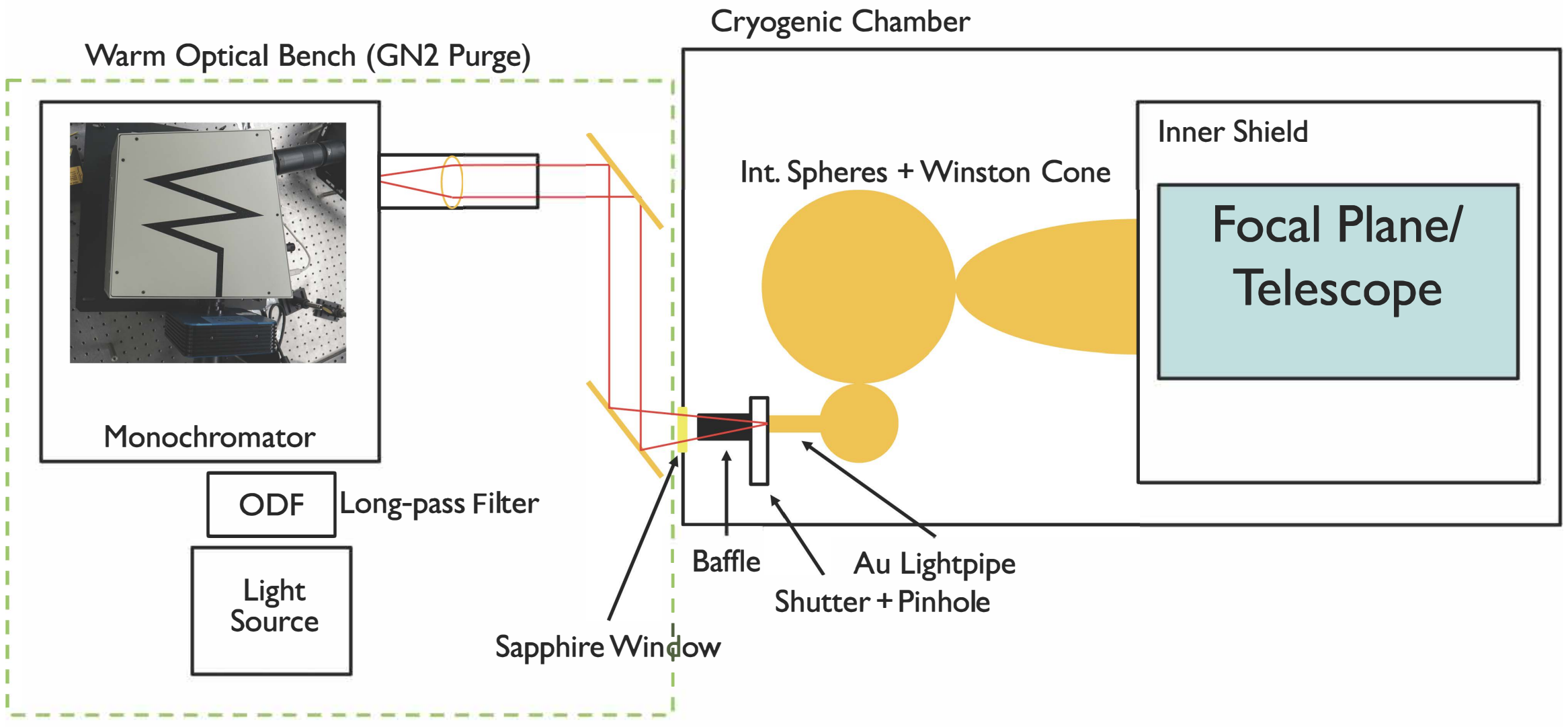

**Figure 3.** Spectral calibration optical configuration. A broadband, tungsten halogen lamp feeds a monochromator on the warm optical bench to generate monochromatic light, which is collimated and directed into the cryogenic chamber. Inside the chamber, two integrating spheres and a cryogenic Winston cone produce a uniform illumination pattern over the field of view of the focal plane or telescope. This configuration is used for both focal plane and system-level calibration.

**Table 1**
SPHEREx Spectral Design Parameters

| | Band 1 | Band 2 | Band 3 | Band 4 | Band 5 | Band 6 |
|---|---|---|---|---|---|---|
| Band Center $\lambda_c$ ($\mu$m) | 0.75–1.11 | 1.11–1.64 | 1.64–2.42 | 2.42–3.82 | 3.82–4.42 | 4.42–5.00 |
| Resolving Power $R = \lambda_c/\Delta\lambda$ | 41.4 | 41.4 | 41.4 | 35 | 110 | 125 |

where the bandwidth is

$$\Delta\lambda = \int \overline{T}_\lambda \, d\lambda, \tag{5}$$

with $\overline{T}_\lambda$ denoting the peak-normalized spectral response of the pixel.

## 3. Spectral Calibration

The SPHEREx spectral calibration campaign generated the data used to characterize the full spectral response function of each optically active pixel across the six detector bands, spanning wavelengths from 0.75 to 5.0 $\mu$m. Because each pixel samples a unique location on the LVF, its effective wavelength response reflects the combined behavior of the telescope optics, dichroic beam splitter, LVFs, and the H2RG detectors. This response varies continuously across the focal plane. Accurate knowledge of the per-pixel spectral function at operating temperature is required to reconstruct sky spectra.

### 3.1. Measurement Overview

The spectral calibration was carried out in two phases that share a common optical architecture but use different cryogenic chambers. The first phase, conducted in late 2022, measured the stand-alone SWIR and MWIR focal plane assemblies using a compact cryogenic chamber developed specifically for SPHEREx at Caltech. In addition to spectral calibration, this facility supported detector readout development and thermomechanical qualification. Although the telescope and dichroic beam splitter were not present in this configuration, the measurements provided early verification of LVF behavior, detector performance, calibration hardware, and the analysis pipeline. This configuration was designed to reproduce the illumination geometry seen by the detectors in flight and to enable early validation of LVF performance and detector response. However, in the absence of the telescope, the LVFs are illuminated by a converging beam rather than the full telescope optical train, resulting in an incidence-angle distribution that differs from the integrated instrument. This difference leads to small but systematic shifts in the effective wavelength solution across the focal plane. As a result, while the focal plane assembly measurements are suitable for validating LVF behavior and analysis methods, they cannot be used to establish the final spectral calibration, which requires the end-to-end telescope configuration.

The second phase, carried out in 2023 November, performed an end-to-end spectral calibration of the fully assembled instrument, providing the complete dataset presented in Section 6. In this configuration, the entire optical train, including the telescope mirrors, dichroic beam splitter, LVFs, and both focal plane assemblies, was installed inside a cryogenic chamber constructed by the Korea Astronomy and Space Science Institute (KASI). This setup reproduces the optical and thermal environment experienced by the instrument in flight.

Figure 3 provides an overview of the spectral calibration system used in both phases. On the warm optical bench, a broadband source feeds a monochromator whose output is focused onto a pinhole aperture inside the cryogenic chamber. The beam then passes through two integrating spheres and a Winston cone (R. Winston 1970) that together produce a uniform illumination pattern over the field of view of the focal plane or telescope.

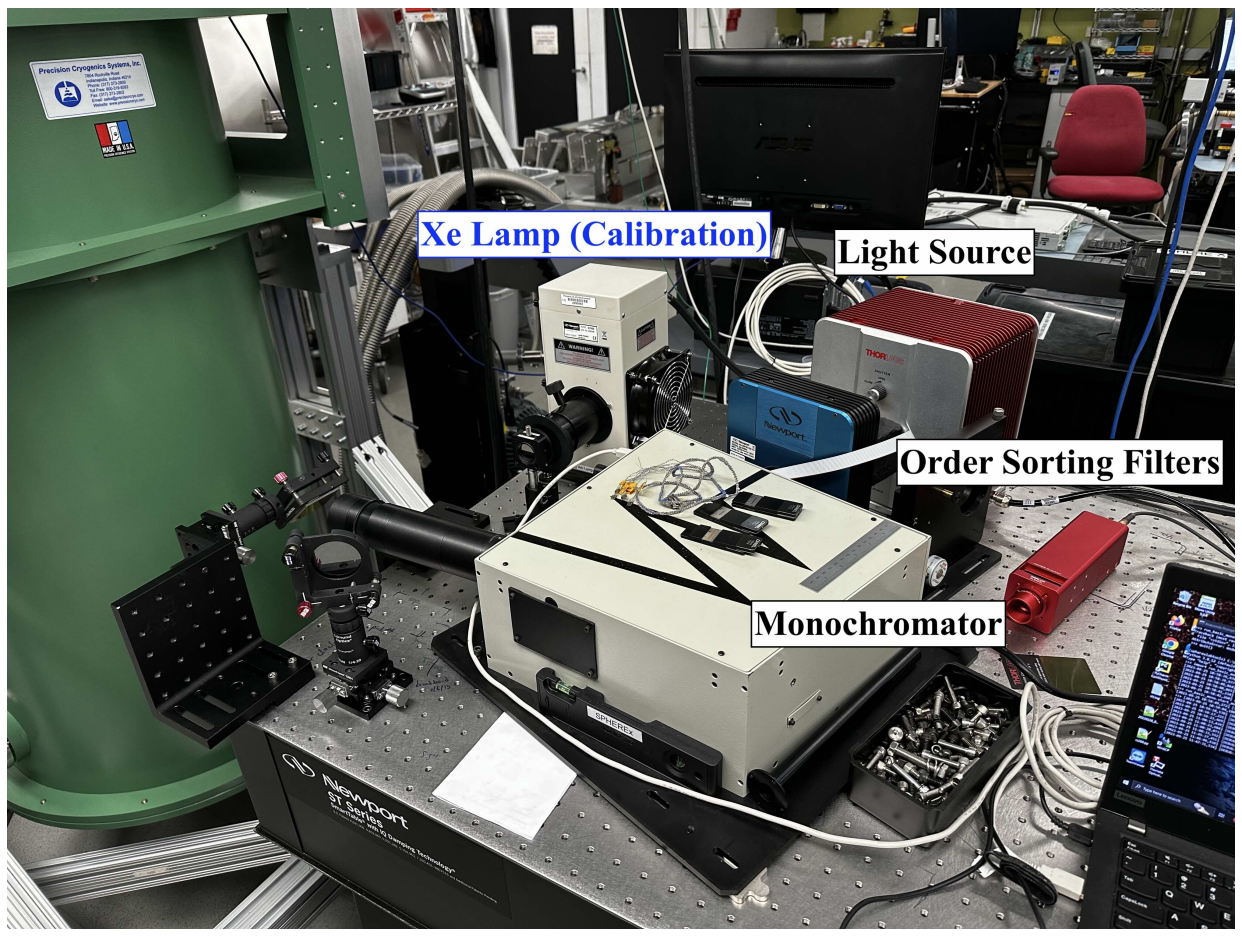

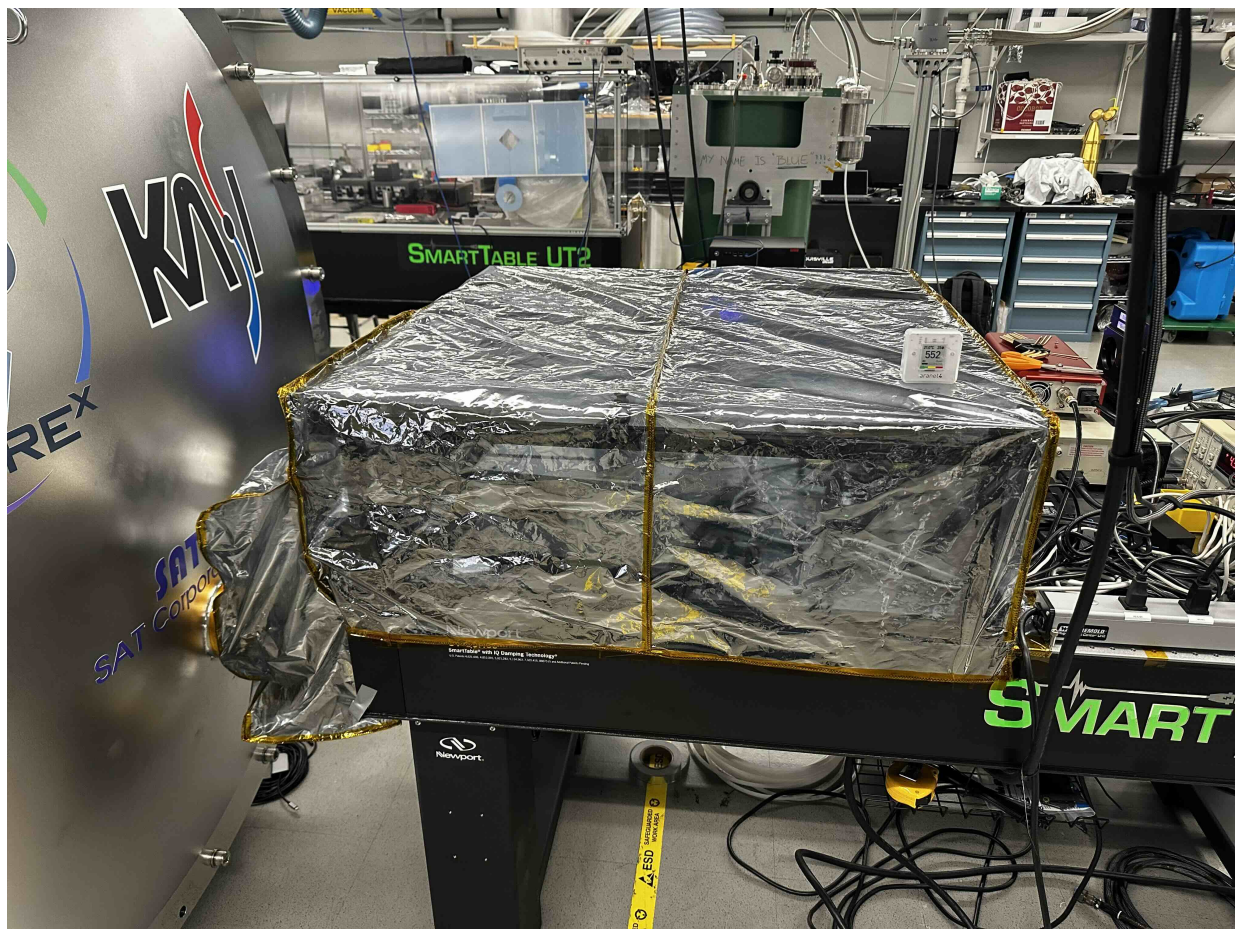

**Figure 4.** Warm optical bench setup. Left: optical bench used during focal plane assembly testing, consisting of a broadband source, monochromator, order-sorting filters, and relay optics. Right: the same bench in operation during full telescope calibration in the KASI chamber, shown with the nitrogen purge system that suppresses atmospheric absorption.

In both phases, the instrument operates at cryogenic temperature and is illuminated with spatially uniform monochromatic light from a monochromator located on the warm optical bench outside the chamber. The wavelength is stepped across the full range to measure the spectral response of each pixel.

The monochromatic beam is collimated and focused by a series of warm mirrors into an $f/5$ beam and then directed through a cryogenic pinhole, with all subsequent optics held at cryogenic temperature so that the background signal remains below the detector dark current ($\sim 10^{-3}\, e^{-}\, \mathrm{s}^{-1}$). At the same time, the room temperature background must be carefully controlled. The optical design restricts the light path such that only room temperature background along with the monochromatic beam is coupled into the cryogenic chamber, while all other background paths are terminated at the pinhole. Even with this design, the signal-to-background ratio typically ranges from 0.1 to 1.5 depending on wavelength, requiring a stable room temperature background for accurate background subtraction.

A traditional approach would use a large cryogenic integrating sphere placed directly in front of the telescope to generate a uniform illumination field (K. Tsumura et al. 2013). For SPHEREx, the large instantaneous field of view would require an integrating sphere nearly 4 m in diameter cooled below 90 K to suppress thermal background. Such a large sphere is impractical, so a more compact system was developed that couples a smaller integrating sphere to a Winston cone. Light entering the Winston cone entrance aperture is redistributed into a controlled output focal ratio without forming an image, producing a uniform illumination pattern across the telescope field of view.

A cryogenic shutter is installed inside the chamber to obtain dark exposures without modifying the optical configuration or altering the thermal environment. Warm optical components outside the chamber are baffled and filtered to block out-of-band light, while the monochromator and order-sorting filters define a narrowband beam with a high signal-to-noise ratio across the full wavelength range.

### 3.2. Warm Optical Setup

Both phases of the calibration campaign use the same warm optical configuration as shown in Figure 4. The bench includes a broadband light source, a monochromator equipped with three diffraction gratings that span the full SPHEREx wavelength range, a set of long-pass order-sorting filters, and relay and focusing optics that send the beam into the cryogenic chamber.

The entire optical path is continuously purged with dry nitrogen gas. The purge suppresses atmospheric absorption features, most notably the water vapor band near 3 $\mu$m and the $CO_2$ band near 4.2 $\mu$m, which would introduce systematic errors into the wavelength calibration.

Further details of the light sources, monochromator configuration, order-sorting filters, and gas purge system are given in the Appendix.

### 3.3. Cryogenic Optical Setup and Thermal Background

The cryogenic optical setup has the shutter, integrating spheres, and Winston cone within the same vacuum chamber as the instrument under test. The optical components are held at $\sim$60 K to suppress thermal background. Although these components are not actively temperature controlled, their temperatures remain within 0.5 K during the measurements.

Monochromatic light generated on the warm optical bench enters the chamber through a sapphire vacuum window, passes through a baffle, and is focused onto a pinhole located inside the cryogenic shutter assembly. With the shutter open, the beam propagates sequentially through two integrating spheres that homogenize the spatial intensity distribution before entering the Winston cone. The Winston cone reshapes the diffuse integrating sphere output into a wide angle beam matched to the SPHEREx field of view. Figures 5 and 6 show the setups for the focal plane and full instrument testing, respectively.

The uniformity of the illumination produced by the Winston cone was measured at the cone output and is shown in Figure 7. This measurement corresponds to the Winston cone used for focal plane level testing, where the cone is placed directly over the focal plane. The focal plane has an optically active area of approximately 12 $\times$ 4 cm, which is fully enclosed within the Winston cone output aperture. For telescope-level testing, a larger Winston cone (shown in Figure 6) is used to illuminate the full telescope aperture. Although a direct uniformity measurement was not performed for the larger cone, it was fabricated using the same design and manufacturing process and is therefore expected to have a similar illumination pattern. A flat-field measurement from this spectral calibration represents a convolution of the Winston cone illumination pattern and the telescope flat field. This is an acceptable intermediate data product, as the final flat-field product is derived from on-sky observations.

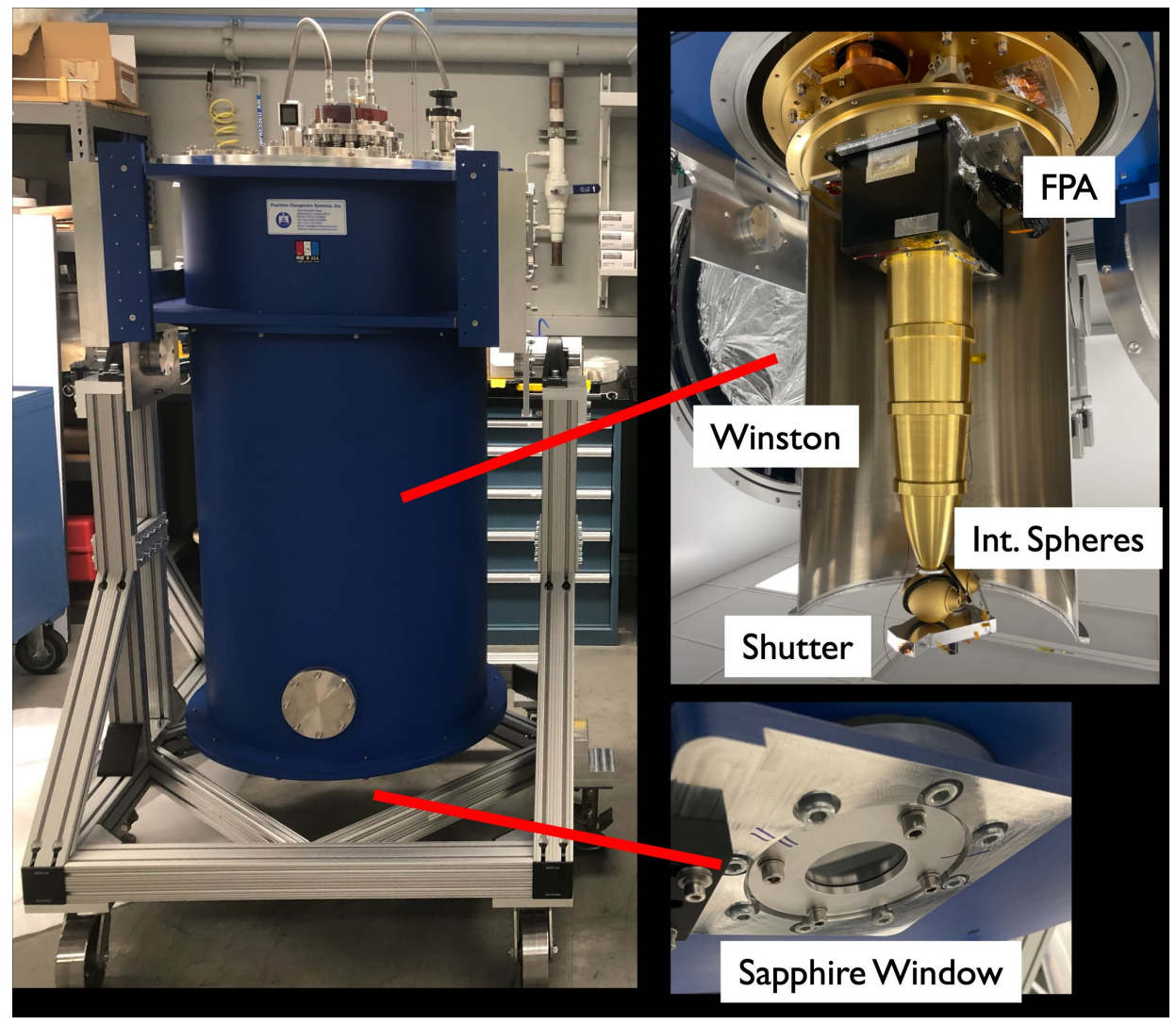

**Figure 5.** Cryostat used for component-level testing of the focal plane assemblies. The focal plane array (FPA), calibration optics, and optically blackened enclosure are mounted on a 20 K cold plate, which is actively temperature controlled to achieve <1 mK stability.

We estimated the power incident on the detectors from 300 K background radiation entering through the sapphire window, propagating through the entrance pinhole of the integrating sphere, and coupling into the Winston cone. The full derivation is provided in Appendix A.2.3, and Figure 8 compares the predicted photocurrent for each band with the measured values. The measurements are consistent with the model, indicating that little unaccounted room temperature stray light is coupled into the cold optics. The background in Bands 1 to 3 remains below the dark current level while the background in Bands 4 to 6 agrees with expectations based on the optical model.

Additional engineering details for the cryogenic chambers, shutter assembly, integrating spheres, and Winston cone, including dimensions and optical modeling, are provided in Appendix A.2.

### 3.4. Control System

The monochromator and warm optical hardware are controlled through the SPHERExLabTools (SLT) software package (S. Condon et al. 2022). SLT provides an interface between the monochromator electronics and the SPHEREx detector readout system, ensuring that wavelength tuning and detector exposures remain synchronized throughout the calibration sequence.

For each spectral scan, SLT commands the monochromator to the desired wavelength, verifies the grating configuration and slit settings, and initiates the associated detector exposure. The system manages exposure timing, metadata logging, housekeeping telemetry, and error checking, and it coordinates the operation of auxiliary hardware such as the shutter and the filter wheel.

SLT is used in both the component-level and system-level spectral calibration campaigns, providing a uniform control environment for all measurements. The same control framework was also used in the focus calibration campaign (S. S. Condon et al. 2024), allowing consistent metadata handling and instrument configuration across the full suite of SPHEREx ground tests.

### 3.5. Measurement Procedure

The primary spectral calibration measurements are obtained by stepping the monochromator through the wavelength range of each band in discrete increments. The wavelength step size for the in-band scans is chosen to be approximately $\Delta\lambda/10$ for each band, which oversamples the spectral response relative to the Nyquist criterion and ensures that the full spectral response function is well resolved. Table 2 summarizes the step size, integration time, and optical configuration used for each band during these scans.

For every spectral band, measurements are acquired in both forward and reverse wavelength order to check for hysteresis and temporal drifts in the optical or thermal system. At every fifth monochromator step, a dark frame is recorded by closing the monochromator shutter. These dark exposures are later subtracted from the corresponding light frames to remove the room temperature background.

Out-of-band scans are also performed to characterize spectral leakage. These scans use a coarse 50 nm wavelength step to probe subpercent-level leakage. The exposure times for the out-of-band scans are fixed at 240 s for Bands 1–4 and 600 s for Bands 5 and 6, whereas the in-band exposure times are scaled with the source brightness at each wavelength relative to the 300 K background. For the SWIR arrays, the in-band measurements span 0.5–2.5 $\mu$m, while for the MWIR arrays they span 2.5–5.2 $\mu$m. The 2.5 $\mu$m transition corresponds to both the intrinsic cutoff of the SWIR detectors and the cutoff of the dichroic beam splitter.

All primary calibration measurements were conducted at the expected in-flight operating temperatures of 62 K for the SWIR detectors and 45 K for the MWIR detectors. To quantify the thermal sensitivity of the spectral response, additional spot-check measurements were collected at the minimum and maximum allowable operating temperatures. For the MWIR arrays these temperatures are 38 K and 57 K, and for the SWIR arrays they are 50 K and 82 K. At each temperature, five wavelength points per band were measured and used to derive a model for temperature-induced spectral shifts.

All scan sequences were executed automatically through the control system described in Section 3.4. Trained operators remained in the laboratory throughout the campaign to monitor system health and verify proper execution of each step as required by JPL flight hardware policy.

In addition to the main calibration scans, several short-duration experiments were performed to assess potential systematics in the illumination and detector response. These tests are described in Section 5.

## 4. Analysis

Each calibration exposure records the detector response to illumination generated by the monochromator. The analysis pipeline transforms these raw detector exposures into fully characterized spectral response functions for every pixel in the SPHEREx focal plane. Processing begins with background subtraction at each wavelength step using the average of the leading and trailing dark frames, followed by normalization to correct for spectral variations in the light sources. For Bands 5 and 6, a deconvolution step removes the effect of the wider monochromator slits required for an adequate signal-to-noise ratio. The in-band and out-of-band measurements are then combined into a unified spectral cube for each pixel, from which the band center and bandwidth are derived. Figure 9 summarizes the complete processing flow from individual exposures to the final per-pixel spectral parameters.

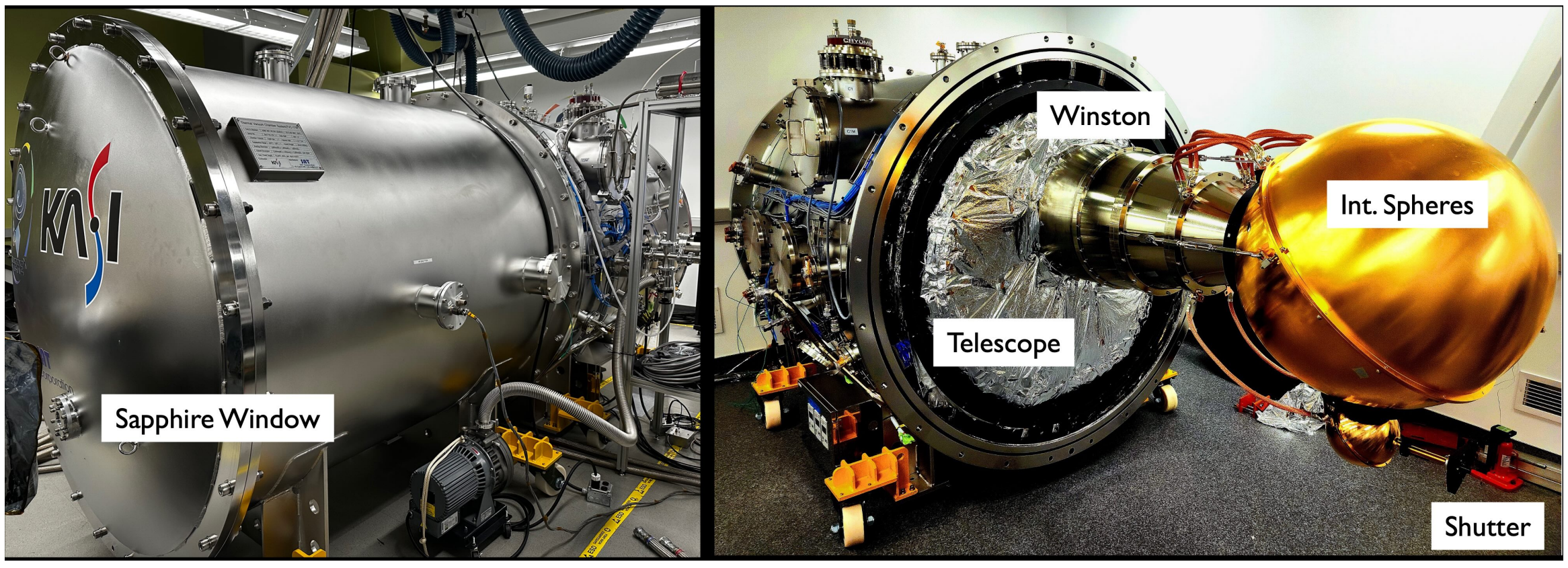

**Figure 6.** KASI cryogenic chamber used for system-level spectral calibration. Left: fully assembled chamber with the monochromatic beam injected through the sapphire window at the front. Right: front half of the chamber with the 300 K vacuum shell removed, showing the Winston cone, two integrating spheres, and cryogenic shutter. The telescope is installed in the rear half of the chamber.

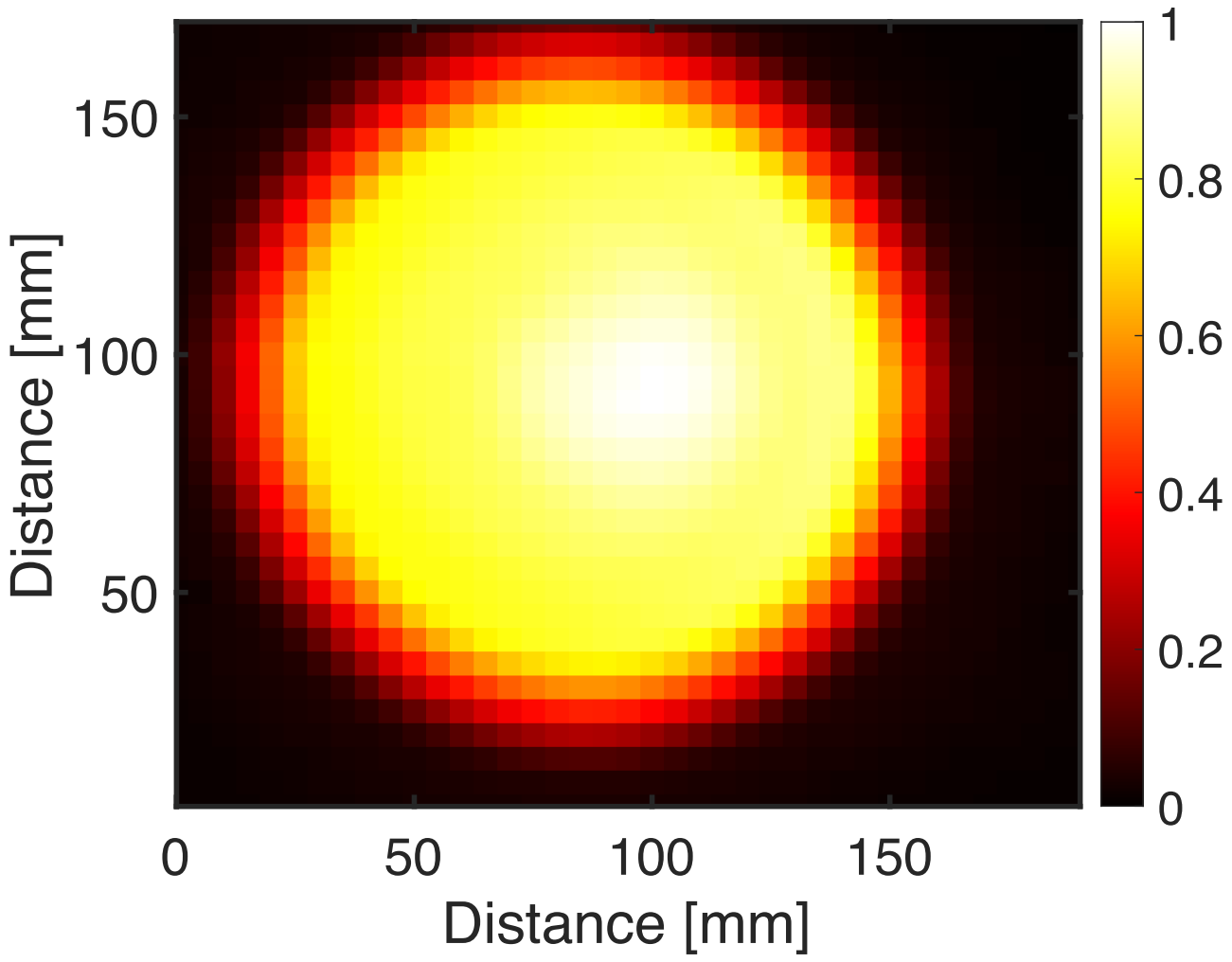

**Figure 7.** Measured illumination pattern at the output of the Winston cone used for focal plane testing. The pattern is largely uniform, with small asymmetries caused by internal baffles in the integrating spheres. Measurements at five wavelengths show similar normalized spatial patterns, indicating that the variations are largely wavelength independent.

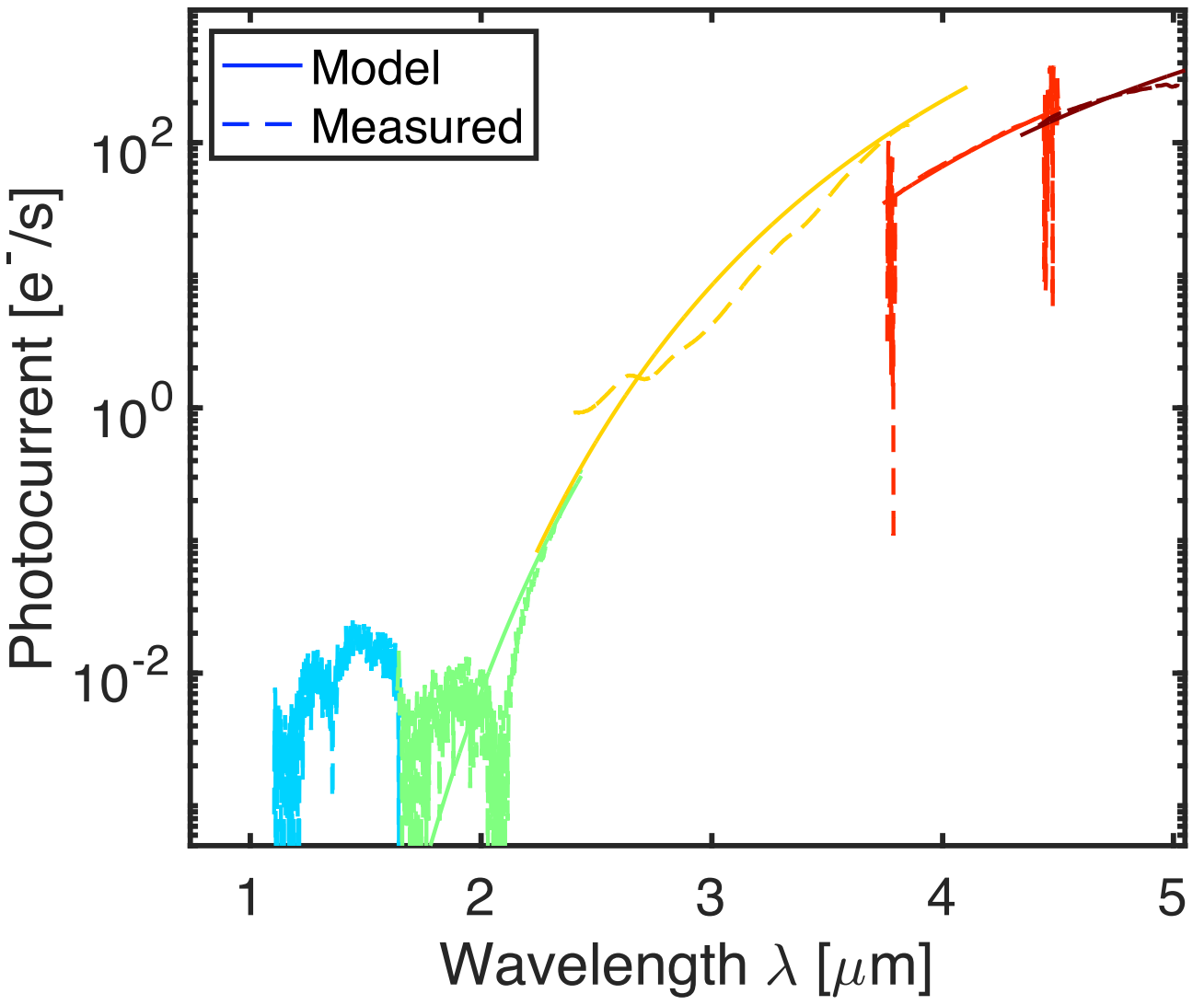

**Figure 8.** Modeled and measured 300 K thermal background loading for the calibration system. Colors denote the individual detector bands, and the same color scheme is used for these bands throughout the paper. Solid curves show the expected photocurrent for each band based on the optical model described in the Appendix, while dotted curves show the measured photocurrent. Band 1 is not shown because its background level lies below the detector dark current. Band 2, shown in dark blue, is consistent with the expected dark current level of approximately 0.01 $e^{-}$ $s^{-1}$. For Bands 3 through 6, the measured photocurrent agrees with the predictions from the optical model. The discontinuities between bands arise from changes in the spectral resolving power $R$.

### *4.1. Background Subtraction*

Background subtraction is performed using dark frames interleaved with the monochromator scan data. Every fifth wavelength step, a dark frame is recorded by closing the monochromator shutter, keeping the optical configuration and detector readout identical to the light exposures. For each light frame, the temporally adjacent dark frame is subtracted to remove the background. This procedure produces a set of background-subtracted images for every wavelength point in the scan.

Leading and trailing dark exposures are differenced to assess residual detector persistence from the light exposures. Persistence signals are found to be lower than the background noise (C. Fazar et al. 2025), so no additional correction is applied.

### *4.2. Exposure Normalization*

After background subtraction, each exposure is normalized to account for the nonflat spectrum of the illumination source and for slow temporal variations in source brightness during

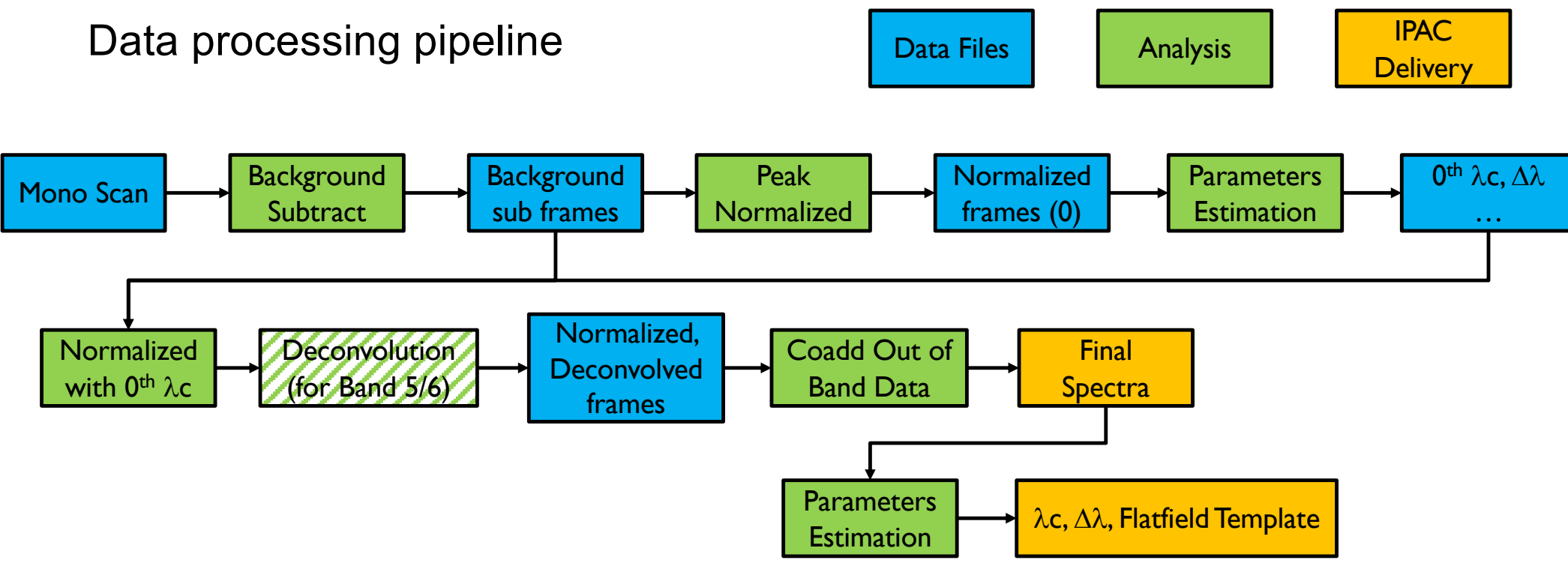

**Figure 9.** Block diagram of the analysis procedure from raw exposures to per-pixel spectral parameters. Blue boxes indicate processed data products, green boxes show the analysis steps, and orange boxes are the delivered calibration products.

**Table 2**
Measurement Setup for the Spectral Calibration Scans

| | Band 1 | Band 2 | Band 3 | Band 4 | Band 5 | Band 6 |
|---|---|---|---|---|---|---|
| In-band Scan ($\mu$m) | 0.675–1.228 | 0.999–1.814 | 1.476–2.676 | 2.280–4.312 | 3.673–4.593 | 4.480–5.100 |
| Step Size ($\mu$m) | 0.002 | 0.003 | 0.004 | 0.004 | 0.003 | 0.004 |
| Grating | 1 | 2 | 2 | 3 | 3 | 3 |
| Order-sorting Filters | 1 | 1 | 2 | 3 | N/A | N/A |
| Slit Size ($\mu$m) | 760 | 500 | 500 | 500 | 760 | 760 |
| Exposure Time (s) | 240 | 50 | 80 | 120 | 280 | 600 |
| Leakage, Short ($\mu$m) | 0.5–0.7 | 0.5–1.0 | 0.5–1.5 | 2.0–2.2 | 2.0–3.7 | 2.0–4.3 |
| Leakage, Long ($\mu$m) | 1.2–2.5 | 1.7–2.5 | N/A | 4.1–5.2 | 4.5–5.2 | N/A |
| Step Size ($\mu$m) | 0.05 | 0.05 | 0.05 | 0.05 | 0.05 | 0.05 |
| Slit Size ($\mu$m) | 760 | 500 | 500 | 500 | 760 | 760 |
| Exposure Time (s) | 240 | 240 | 240 | 240 | 600 | 600 |

**Note.** The top panel shows the configuration used for in-band measurements, consisting of finely stepped exposures that cover the main bandpass of each channel. The bottom panel shows the configuration used for out-of-band measurements, which use a coarser wavelength steps to characterize out-of-band leakage. The combination of grating dispersion and slit size sets the monochromatic beamwidth to approximately 10% of $\Delta\lambda$. Further details of the monochromator model and configuration are provided in Appendix A.1.2

the scan. The illumination follows an effective 3400 K blackbody spectrum, with additional spectral features introduced by the lamp filament, vacuum windows, and optical coatings. A secondary power meter was initially installed on the warm optical bench to monitor source brightness using a pickoff mirror to divert approximately 1% of the total optical power, but the power meter was not sufficiently sensitive. An empirical normalization method was therefore adopted.

The normalization factor is derived iteratively. A first estimate is obtained from the factory-provided spectral profile of the lamp and monochromator. Following an initial extraction of band centers for all pixels, the median detector signal within each 2 nm wavelength bin is computed and used to update the normalization function. A new set of band centers is then derived from the renormalized data, and this procedure is repeated until convergence is achieved across the full wavelength range. For monochromator settings outside the primary SPHEREx bandpass, the normalization factor is interpolated using a model of the monochromator and illumination source, together with measured normalization values near the band edges. Figure 10 shows the resulting normalization spectrum used in the analysis.

### 4.3. Spectral Deconvolution

In the calibration setup summarized in Table 2, the monochromator slit widths are chosen such that the output bandwidth of the monochromatic illumination is approximately an order of magnitude narrower than the design resolving power $R$ at the corresponding wavelength. For example, in Band 1 at approximately 1 $\mu$m with $R \sim 41$, corresponding to $\Delta\lambda \sim 24$ nm, the first grating with a dispersion of 3.1 nm mm$^{-1}$ is used in combination with 760 $\mu$m slits, resulting in a monochromator bandwidth of approximately 2.4 nm. For Bands 5 and 6, the monochromator operates with the third grating, which has a dispersion of 25.8 nm mm$^{-1}$. At these longer wavelengths, sufficient optical throughput requires the use of 760 $\mu$m slits, yielding a monochromator bandwidth that is only a factor of 2 narrower than the intrinsic SPHEREx bandpass. Recovering the intrinsic SPHEREx spectral response therefore requires deconvolving the measured spectra with the monochromator slit function.

The observed spectrum $I_{\rm obs}(\lambda)$ is the convolution of the intrinsic response $I_{\rm int}(\lambda)$ and the monochromator slit function $S(\lambda)$:

$$I_{\rm obs}(\lambda) = (I_{\rm int} * S)(\lambda). \qquad (6)$$

The intrinsic response $I_{\rm int}(\lambda)$ is recovered using the Lucy–Richardson iterative deconvolution algorithm (L. B. Lucy 1974). A subset of measurements is repeated with 200 $\mu$m slits and tenfold longer integration times to independently verify the deconvolution accuracy. The nominal slit function for a monochromator with rectangular slits is a top-hat profile whose width matches the physical slit width. However, using a pure top-hat kernel produces noticeable ringing in the deconvolved spectra. To mitigate this effect, and to better match the measured line profiles obtained with 200 $\mu$m slits, we adopt a slightly tapered slit function that smoothly transitions to zero at the edges. This modified kernel suppresses ringing artifacts and improves agreement with validation data. Because the 200 $\mu$m measurements are nearly an order of magnitude narrower than the SPHEREx bandpass, they can be treated as the effective "true" monochromatic illumination for validating the deconvolution.

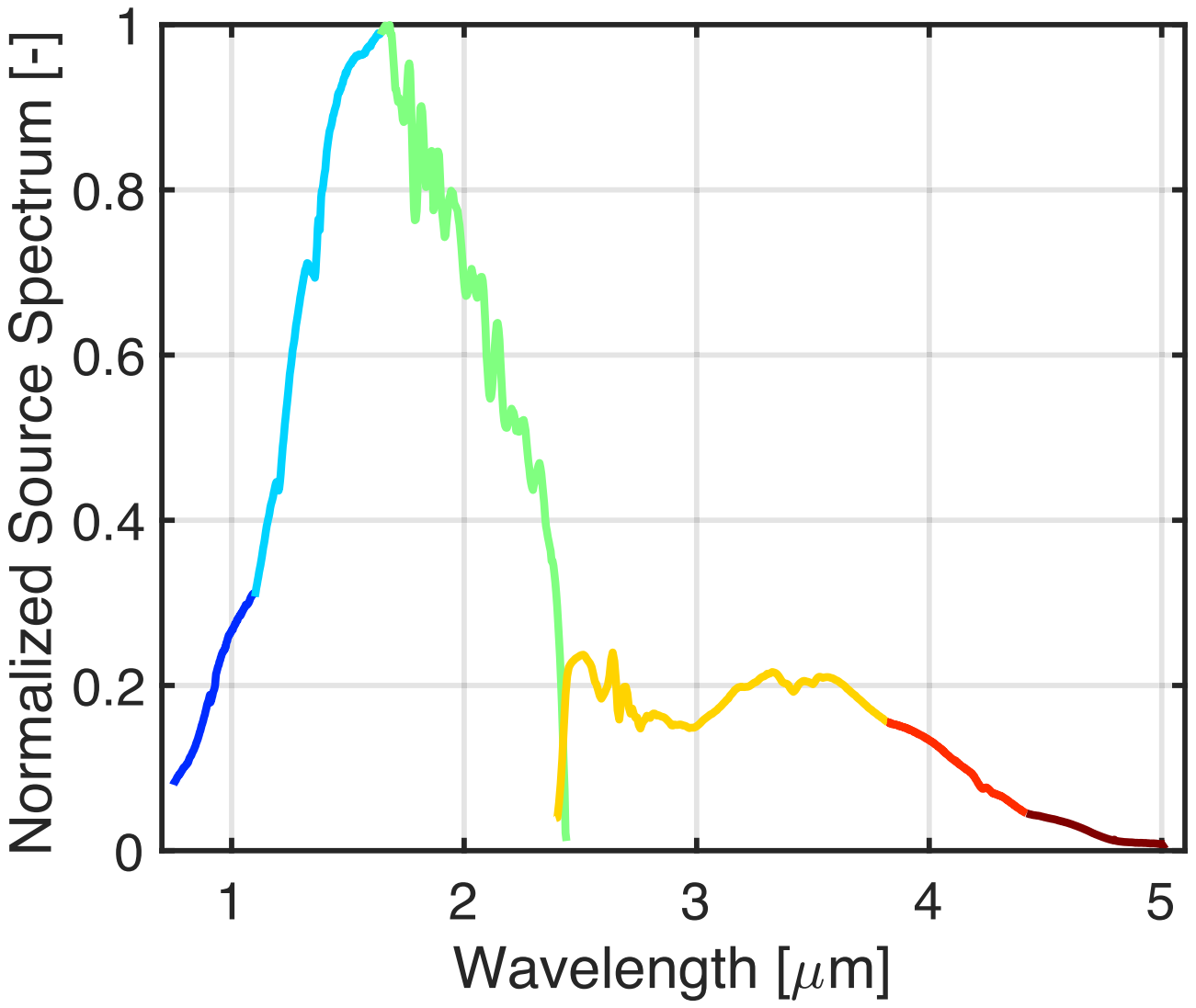

**Figure 10.** Final normalization spectrum used in the analysis pipeline. The curve reflects the combined spectral transmission of the lamp, monochromator, warm optics, sapphire window, integrating spheres, Winston cone, and the intrinsic instrument spectral response.

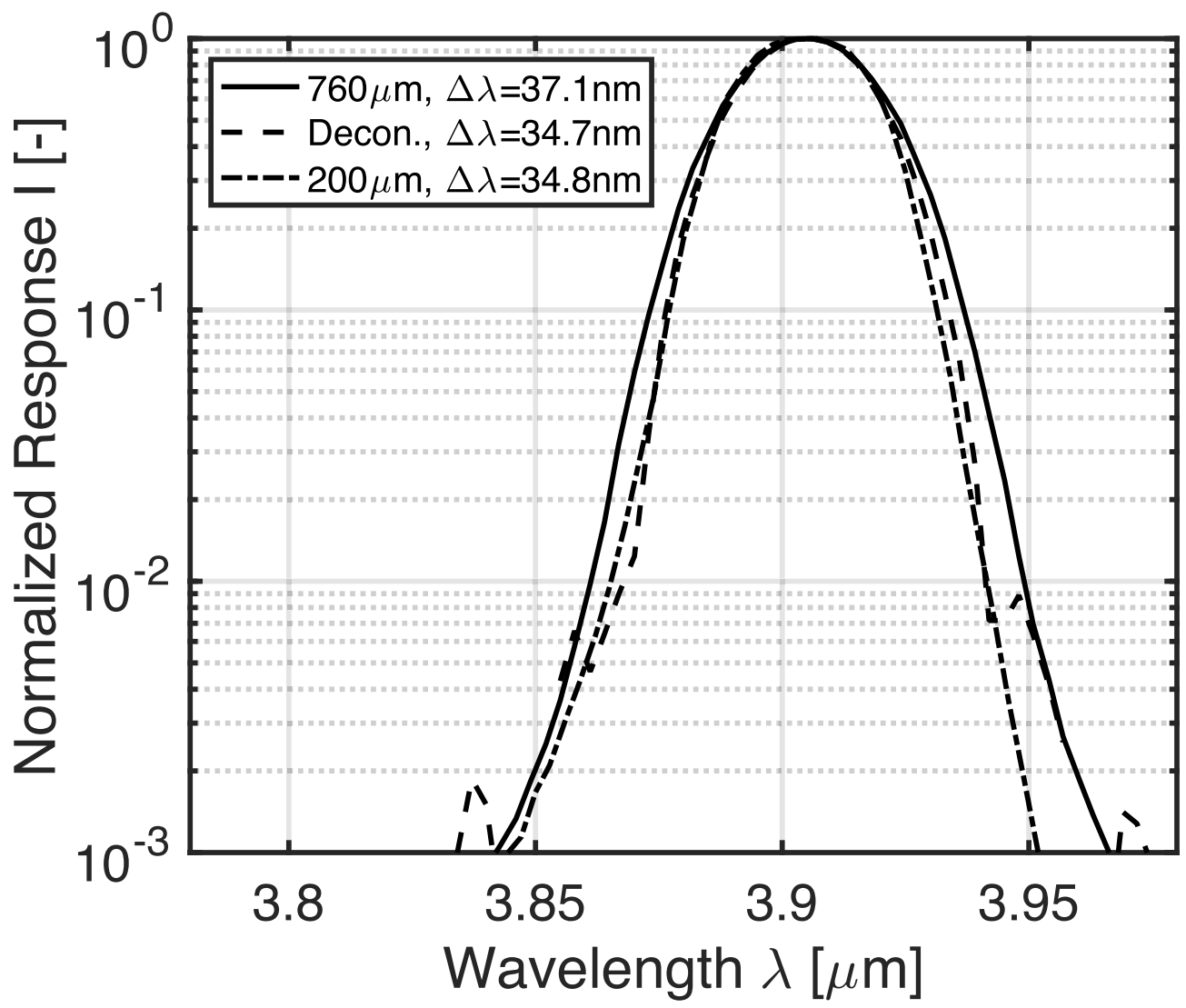

**Figure 11.** Comparison between spectra obtained with narrow slits and spectra recovered through deconvolution at approximately 3.9 $\mu$m. The deconvolved spectrum reproduces the intrinsic bandwidth measured with the 200 $\mu$m slits with approximately 10 times the exposure time.

Figure 11 compares these narrow slit measurements with the deconvolved spectra. The intrinsic bandwidth and line shape are reproduced to within the measurement uncertainties.

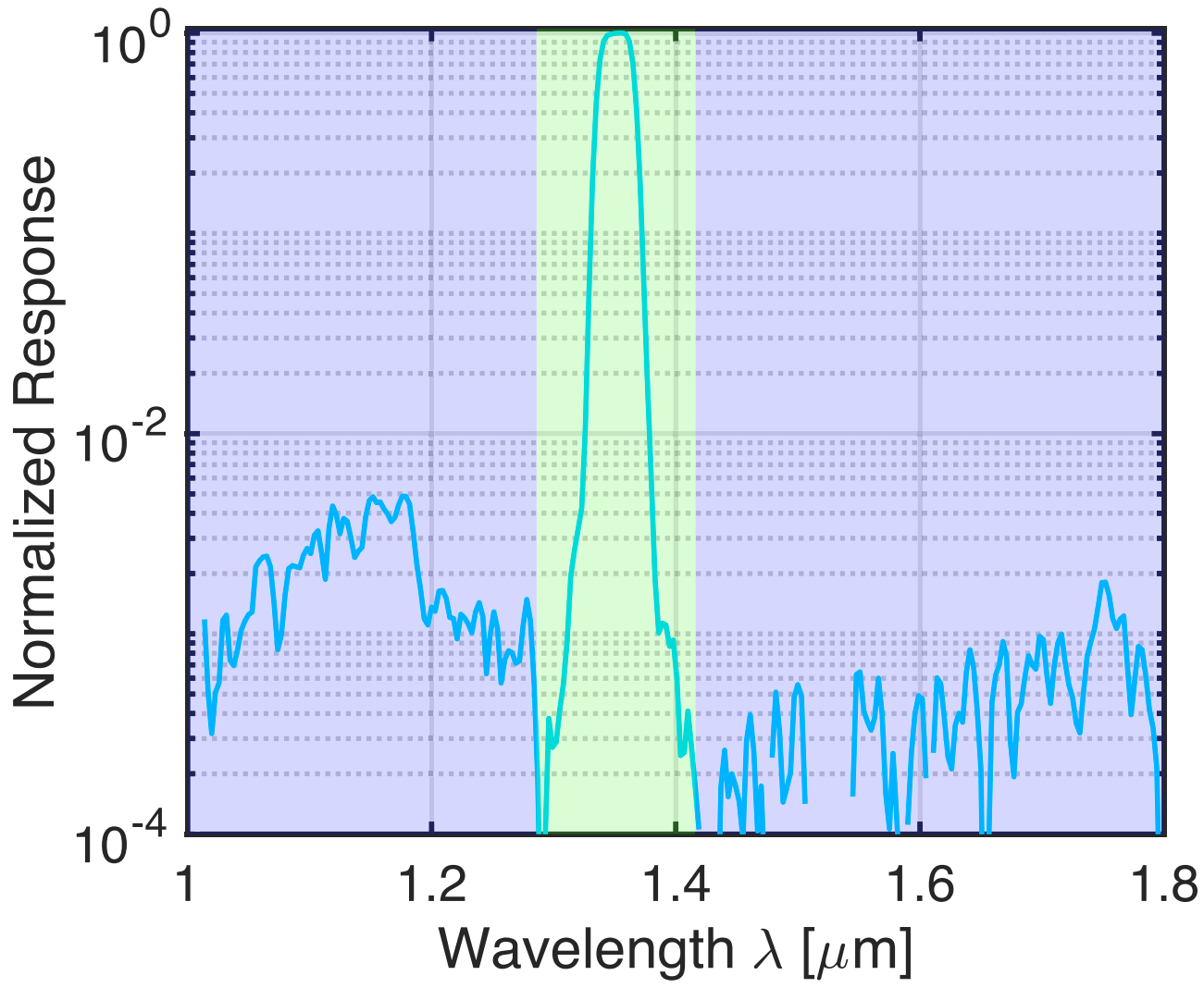

**Figure 12.** Example spectral response for a representative pixel in Band 2. The main response (green) is measured independently for this pixel using the high-resolution in-band scan. The much weaker out-of-band response (purple), which varies only on large spatial scales across the focal plane, is obtained by coadding pixels with similar band-center values to increase the signal-to-noise ratio.

### *4.4. Final Spectra and Parameter Estimation*

For each pixel, the spectral response is constructed by combining the background-subtracted and spectrally normalized high-resolution in-band scan, with a step size of 0.002–0.004 $\mu$m, and the coarser out-of-band scan, with a step size of 0.05 $\mu$m. The in-band response varies at the few percent level between pixels in the same spectral channel, most likely due to per-pixel gain variations that must be measured independently. Out-of-band leakage is observed in Bands 1 and 2 at approximately 1% of the main spectral peak, and is consistent with noise at other wavelengths, as discussed in Section 6.2. The wavelength peak of the leakage in Bands 1 and 2 shifts with the LVF wavelength progression and exhibits similar level of leakage for pixels sharing the same main band wavelength, indicating that the leakage is most likely set by the LVF coating rather than by individual detector pixels. To improve the signal-to-noise ratio in these regions, the out-of-band measurements are coadded in wavelength bins, as illustrated in Figure 12.

Each pixel spectral response is peak normalized using the source spectrum shown in Figure 10. As a result, pixel-to-pixel flat-field variations are not measured in this calibration. The telescope flat field is derived using a separate pipeline based on in-flight zodiacal-light observations (H. Hui et al. 2026, in preparation), and the final absolute photometric gain calibration will rely on observations of well-characterized stellar standards obtained with JWST (M. L. N. Ashby et al. 2026, in preparation).

Per-pixel band centers and bandwidths are computed using the definitions in Section 2. These parameters form the basis of the full spectral calibration data cube described in Section 6.1.

## 5. Measurement Systematics

A series of dedicated measurements was used to validate the derived wavelength calibration, assess systematic effects, and provide absolute spectral references including a characterization of the out-of-band response.

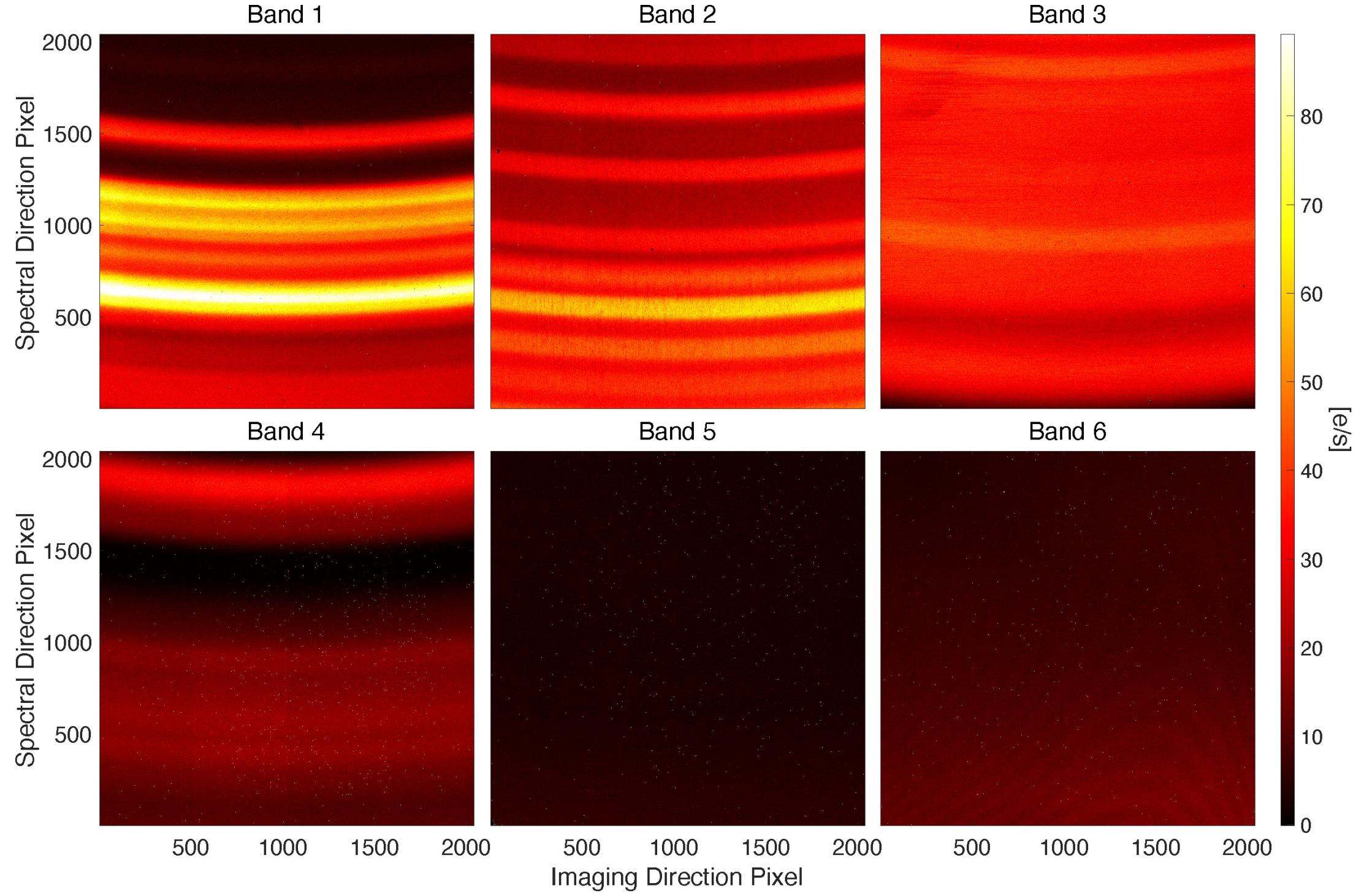

**Figure 13.** Exposure obtained by placing a xenon arc lamp directly in front of the vacuum window. Multiple xenon emission lines are detected across Bands 1 through 3 at the SPHEREx spectral resolution. No identifiable features are present above 2.5 $\mu$m, preventing a direct absolute calibration of the third monochromator grating.

### *5.1. Absolute Wavelength Calibration*

The absolute wavelength calibration depends on the accuracy of the monochromator calibration. The monochromator employs three gratings to cover the range from 0.5 $\mu$m to 5 $\mu$m. Gratings 1 and 2, which are used for Bands 1 through 3 and part of Band 4, are cross-calibrated using the bright emission lines from a high-power xenon arc lamp (Figure 13). One-dimensional spectra extracted from these exposures are matched to reference xenon line lists (C. J. Humphreys 1973), yielding absolute wavelength uncertainties of $\pm 0.6$ nm for grating 1 and $\pm 0.9$ nm for grating 2.

Grating 3, which covers wavelengths above 2.5 $\mu$m, cannot be calibrated directly with the xenon lamp because no measurable emission lines are present above 2.5 $\mu$m. A relative calibration is therefore obtained by comparing monochromator measurements taken with gratings 2 and 3 in their overlap region from 2.4 $\mu$m to 2.6 $\mu$m. However, the overlap lies within Band 4, which has the lowest spectral resolving power in the instrument with $R = 35$ ($\Delta\lambda = 71$ nm at 2.5 $\mu$m). The coarse resolution of Band 4 limits the ability to refine the absolute calibration of grating 3. In addition, the dispersion of grating 3 is 2.7 times larger than that of grating 2, which restricts the precision of this comparison to approximately 5 nm.

An additional cross-check was performed using the atmospheric $CO_2$ absorption feature near 4.2 $\mu$m. However, this feature is relatively broad, and its shape depends on both concentration and pressure (M. F. Modest 2013). As a result, it serves as a useful qualitative validation but does not improve the absolute wavelength precision beyond approximately 5 nm.

### *5.2. Scan Direction Effects*

During the calibration scans, the monochromator steps sequentially through wavelength in increasing and decreasing order. Because each exposure is taken immediately after the previous one, detector persistence, slow temporal drifts in the illumination system, and mechanical hysteresis in the monochromator grating drive could introduce direction-dependent systematics. To evaluate these effects, two complete datasets were acquired for every band, one with wavelengths scanned in ascending order and one in descending order. Comparing these datasets also probes the long-term thermal stability and repeatability of the optical setup, as each full spectral scan requires multiple days to complete.

The derived spectral parameters from the forward and reverse scans are compared on a per-pixel basis. Differences in band center are below 1 nm, and no statistically significant variation is observed in band center (Figure 14).

### *5.3. Slit Alignment and Position*

At the monochromator output, the wavelength varies across the beam due to the finite slit width and the dispersion of the selected grating. If the beam is clipped as it enters the integrating spheres inside the cryogenic chamber, the resulting transmission profile becomes asymmetric, which can introduce a shift in the effective wavelength of the input illumination. This behavior was observed during early commissioning, when small steering offsets led to partial truncation of the beam at the cryogenic pinhole.

In the original design, the pinhole aperture was made small to reject room temperature background loading. In practice, the minimum achievable beamwidth at the pinhole was approximately 5 mm. To mitigate sensitivity to slit alignment, the pinhole aperture in the shutter assembly was increased to 5 mm, ensuring that the full beam footprint was transmitted even in the presence of small angular deviations. In addition, the monochromator output was realigned at the start of each calibration run to ensure that the beam was centered on the pinhole.

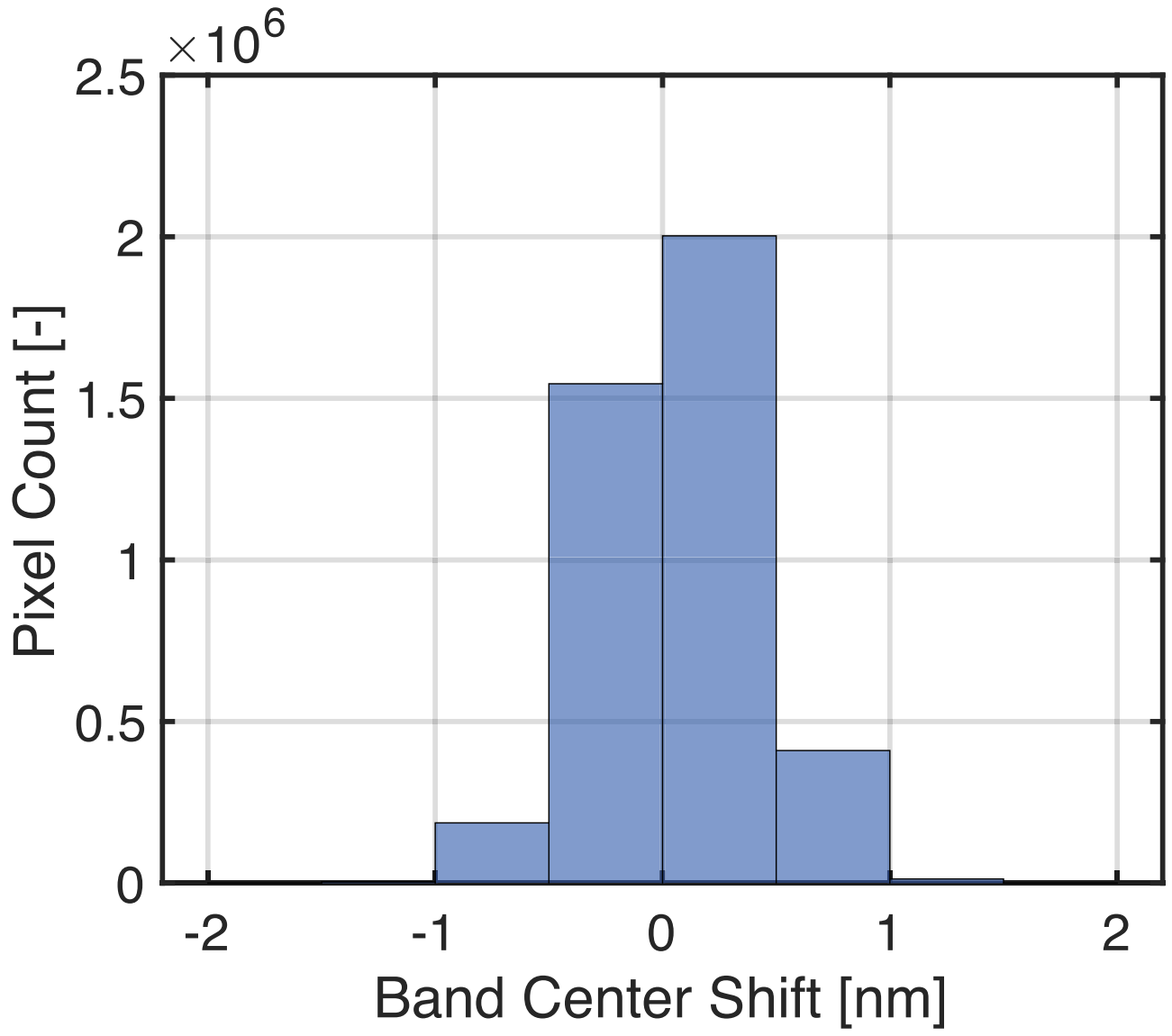

**Figure 14.** Histogram of the difference in band center derived from forward and reverse wavelength scans. The distribution is centered at zero and has a width below ±1 nm, demonstrating that scan direction does not introduce a measurable systematic shift.

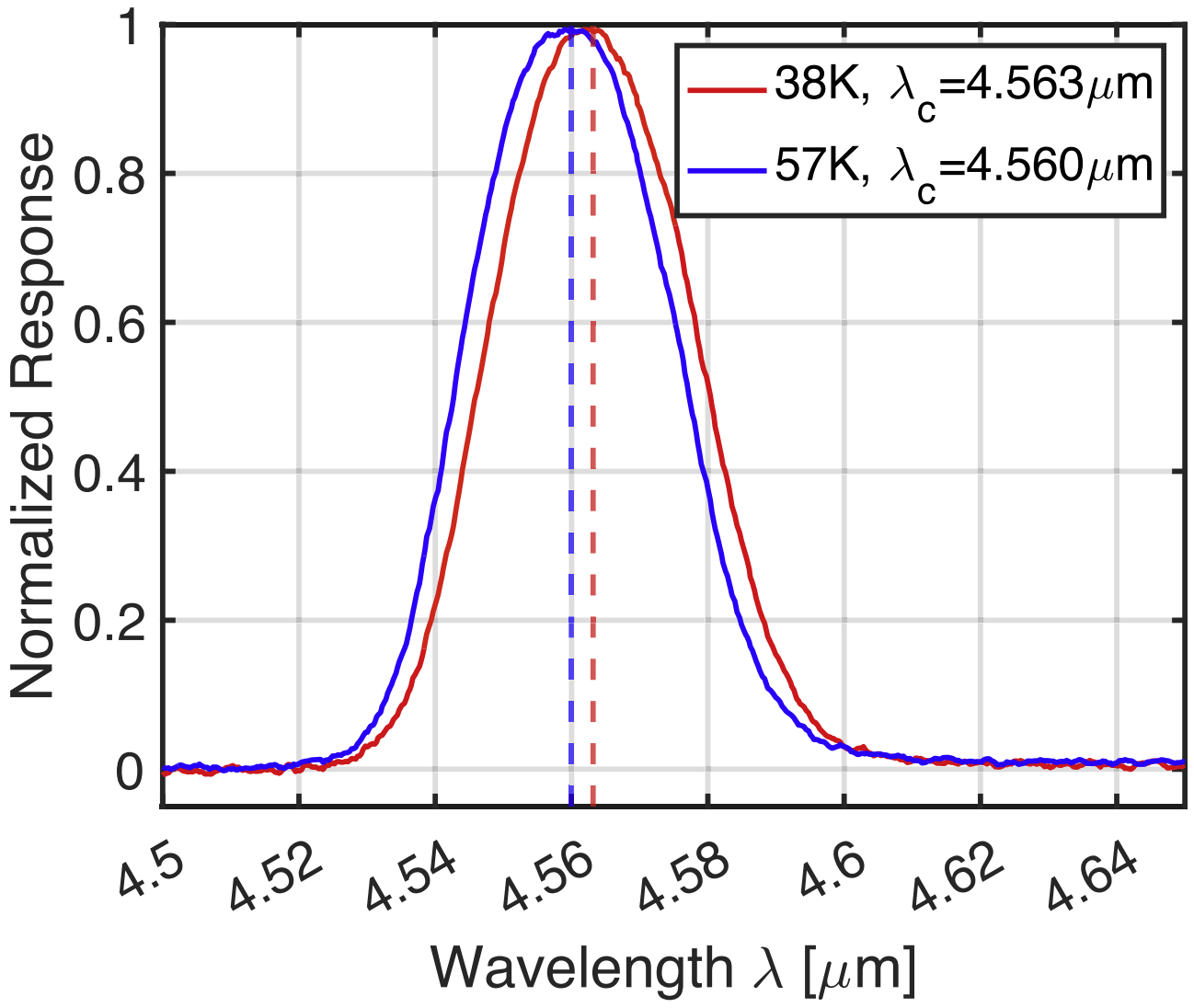

**Figure 15.** Example temperature test for Band 6 (MWIR), showing the band center measured at the minimum and maximum allowable focal plane temperatures of 38 K and 57 K, respectively. The band center shifts by approximately 3 nm over this 19 K range.

A series of dedicated tests was performed to quantify the residual sensitivity to misalignment. During these tests, the beam was intentionally steered by controlled amounts while monitoring the apparent band center of selected detector pixels. The resulting wavelength shifts remained at the level of approximately 0.5 nm, demonstrating that slit position effects contribute negligibly to the overall wavelength uncertainty in the final calibration.

### *5.4. Temperature Dependence*

Most of the calibration scans were acquired at the nominal operating temperatures of 62 K for the SWIR detectors and 45 K for the MWIR detectors. To characterize the thermal sensitivity of the spectral response and to develop a temperature-dependent wavelength model, additional spot-check measurements were taken at the minimum and maximum allowable operating temperatures: 50 K and 82 K for the SWIR detectors, and 38 K and 57 K for the MWIR detectors, respectively. Five wavelength points were sampled in each band at these temperatures to quantify the resulting band-center shifts (Figure 15).

For the SWIR arrays, increasing the focal plane temperature from 48 K to 82 K produces a total band-center shift of approximately 2 nm across the full range. For the MWIR arrays, increasing the temperature from 38 K to 57 K results in a shift of roughly 3 nm.

After launch, both focal plane modules achieved their target temperatures within 0.1 mK using active thermal regulation. The thermal control system has sufficient margin to maintain at these temperatures and stability over the 2 yr mission lifetime (J. J. Bock et al. 2026; P. M. Korngut et al. 2026). Given this level of stability, temperature-induced wavelength variations are negligible in flight, and no temperature correction is applied in nominal science operations.

## 6. Results

The spectral calibration campaign described above resulted in a primary data product consisting of the peak-normalized spectral response function for each of the 25 million optically active pixels across all six detector bands, spanning 0.75–5.0 $\mu$m. For every pixel, the band center $\lambda_c$ and resolving power $R = \lambda_c/\Delta\lambda$ were derived from these measurements. In addition to these per-pixel products, spectral response functions were coadded within defined pixel ensembles to construct fiducial bandpass functions for routine science analysis. These data are part of the SPHEREx data supplement release along with the mission data products (R. Akeson et al. 2025).

### *6.1. Per-pixel Spectral Parameters*

Many SPHEREx science applications require only the measured band center $\lambda_c$ and resolving power for each pixel rather than the full spectral data cube. Figure 16 shows the spatial distribution of the measured band centers across all six detectors, divided into 17 spectral channels per band.

Figure 17 shows the number of functioning pixels in each spectral channel. The distribution is nearly uniform across all channels, indicating that the measured spectral progression closely follows the LVF design. A pileup feature appears at 2.42 $\mu$m, corresponding to the dichroic beam splitter transition between Band 3 and Band 4. Near this transition, the optical efficiency decreases rapidly, producing a progressively narrower and more asymmetric bandpass that causes the inferred band centers of edge pixels to cluster at the transition wavelength.

Although each detector band is designed to achieve a constant resolving power, measurable variation is observed across the focal plane, as shown in Figure 18. These variations track the wavelength progression of the LVF and reflect fabrication nonuniformity. Bands 5 and 6 exhibit fringing patterns caused by small variations in the LVF-to-detector spacing. This effect is detectable only in the longer-wavelength bands, as the lower resolving power in the shorter-wavelength bands smears the signal. We note that flat-field measurements show that the fringe positions are not stable between laboratory and in-flight measurements, suggesting that the corresponding fringing patterns in resolving power may also differ between these environments. Figure 19 shows resolving the power as a function of band center, illustrating that Bands 2–4 are relatively flat, while Bands 1, 5, and 6 show a modest slope but remain within design requirements.

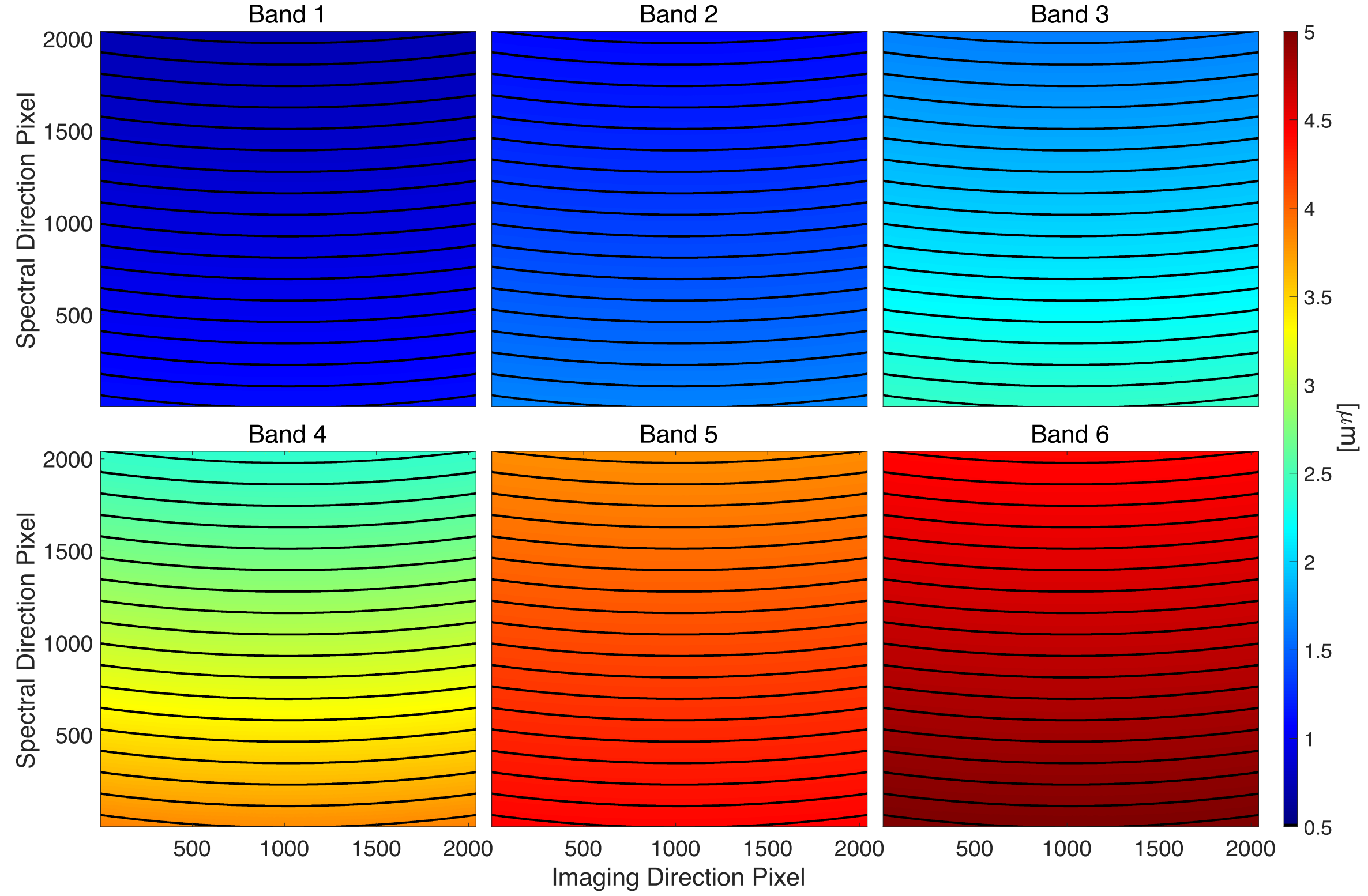

**Figure 16.** Measured band-center maps $\lambda_c$ for all six SPHEREx detectors. Black lines mark the boundaries of the nominal 17 spectral channels.

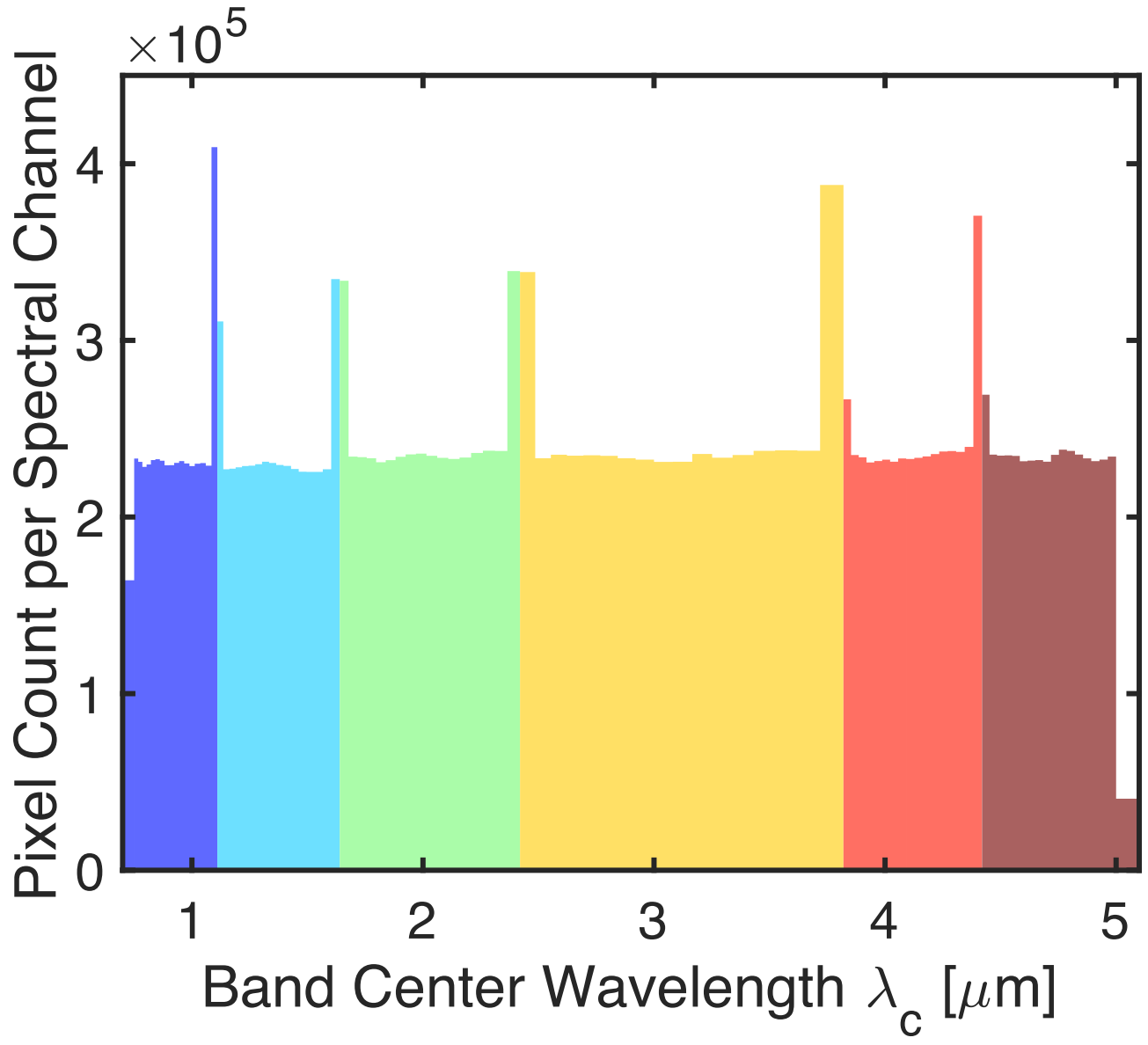

**Figure 17.** Histogram of functioning pixel counts per spectral channel, with different colors corresponding to different detector bands. The horizontal extent of each bin reflects the wavelength coverage of the corresponding channel. The pixel distribution is largely uniform, indicating that the wavelength progression across the focal plane behaves as expected. Additional pixels between detector bands arise from intentional wavelength overlap between adjacent band edges, ensuring continuous wavelength coverage.

A full listing of the median spectral parameters for each channel, including $\lambda_c$ and resolving power, is provided in the SPHEREx Explanatory Supplement (R. Akeson et al. 2025). Pixels with nonphysical spectral parameters are identified through an iterative outlier-rejection procedure applied to both the band-center and bandwidth maps. Most rejected pixels correspond to low quantum efficiency or an anomalously high dark current. Table 3 summarizes the number of flagged pixels in each band. They represent fewer than 0.1% of the optically active pixels and constitute one component of the overall mission bad-pixel mask.

### *6.2. Out-of-band Response and Verification*

The out-of-band spectral response was characterized using the dedicated leakage scans described in Section 3.5. No significant out-of-band leakage was detected in Bands 3 through 6. Bands 1 and 2 exhibit measurable leakage on both the red and blue sides, as illustrated in Figure 20, which shows the detector response to a 1.064 $\mu$m laser. These features track the band-center progression and appear at approximately $0.73\lambda_c$, $0.87\lambda_c$, and $1.3\lambda_c$, with peak amplitudes of about 0.8%, 0.5%, and 0.3%, respectively. When integrated over the full bandpass, the total out-of-band throughput remains below $5 \times 10^{-3}$ and does not significantly impact the effective resolving power or band-center extraction.

The measured out-of-band response was incorporated into the SPHEREx sky simulator to assess its effect on science analysis (B. P. Crill et al. 2025). This contribution is particularly important in Band 1 near the strong 1.083 $\mu$m helium airglow line, where upper-atmosphere emission can exceed the zodiacal-light background by more than an order of magnitude, and is present in every exposure. The mission data processing pipeline includes a calculation to model and subtract this helium airglow emission, including the out-of-band component introduced by the measured leakage.

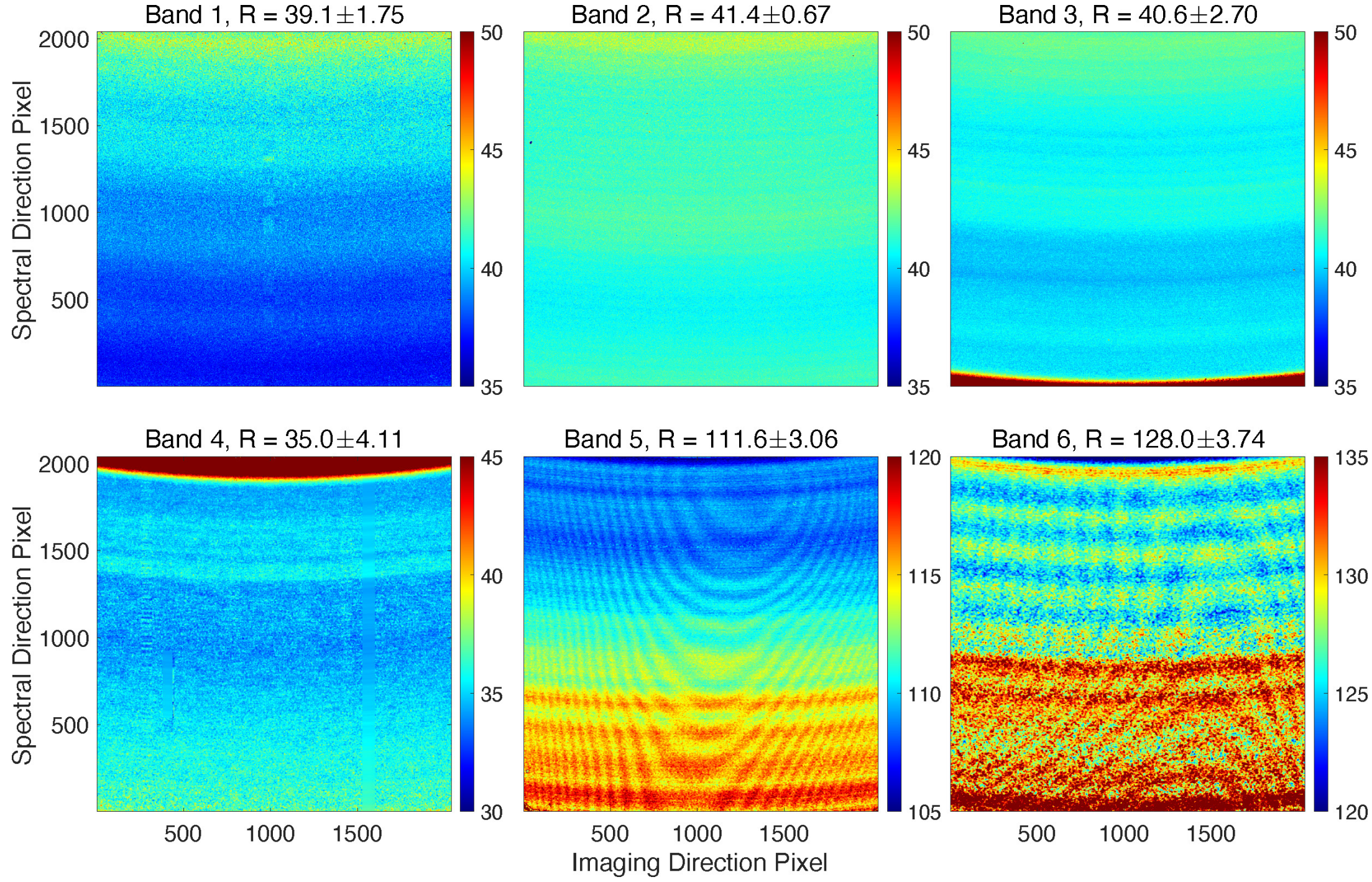

**Figure 18.** Resolving power maps for the six SPHEREx detectors. The spatial variations trace the LVF wavelength gradient and reflect intrinsic LVF fabrication structure. The sharp features at the boundary between Bands 3 and 4 are real and are caused by the dichroic beam splitter transition. Fringing in Bands 5 and 6 arises from small variations in the LVF-to-detector spacing. The fringes are more apparent at higher resolving power.

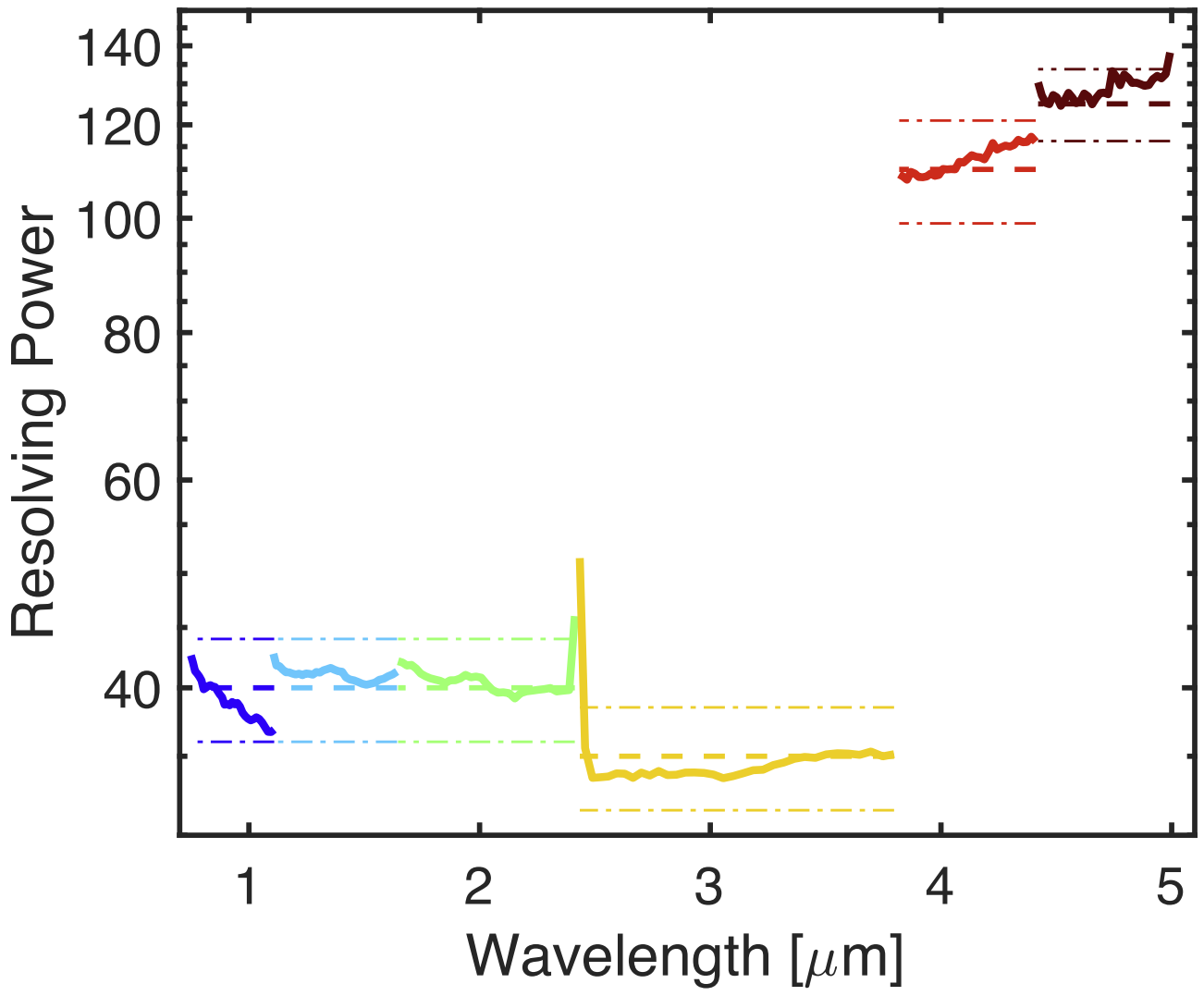

**Figure 19.** Resolving power as a function of band center for all detectors. Dashed lines indicate the target resolving power. The ±10% requirement for Bands 1–5 and ±7% requirement for Band 6 are shown as dotted–dashed lines. Bands 2 to 4 show flat trends, while Bands 1, 5, and 6 show mild slopes but remain within design tolerance. The peak at 2.42 $\mu$m is produced by the dichroic beam splitter transition.

Narrow-line laser sources were used to provide an independent verification of the out-of-band measurements. A 1.064 $\mu$m laser (calibrated to ±2 nm) and a 1.548 $\mu$m laser (calibrated to ±10 nm) were injected into the illumination path. Both measurements reproduced the same out-of-band features observed in the broadband scans, confirming that these signals originate from the instrument response rather than from calibration systematics (Figure 21).

**Table 3**
Number of Pixels Flagged for Nonphysical Spectral Parameters in Each Band

| Band | Flagged Pixels | Fraction (%) |
|---|---|---|
| 1 | 1697 | 0.040 |
| 2 | 1774 | 0.042 |
| 3 | 1780 | 0.042 |
| 4 | 2790 | 0.067 |
| 5 | 1813 | 0.043 |
| 6 | 1599 | 0.038 |

### *6.3. Fiducial Spectral Response Functions*

For routine use, a set of fiducial spectral response functions was constructed to represent the characteristic behavior of groups of pixels within each spectral channel.

Two definitions of the fiducial response are provided. The first averages the spectra of pixels located near the geometric center of each nominal channel. This method preserves the intrinsic channel bandwidth and yields a band center corresponding to the local midpoint of the LVF rather than the mean over the full pixel ensemble.

The second method averages the spectra of all pixels assigned to the same spectral channel. This produces a spectrum representative of the full pixel population and results in a modest broadening of the effective resolving power. This definition better reflects the response encountered in the survey data, where each astrophysical source is observed multiple times at different locations within a given channel.

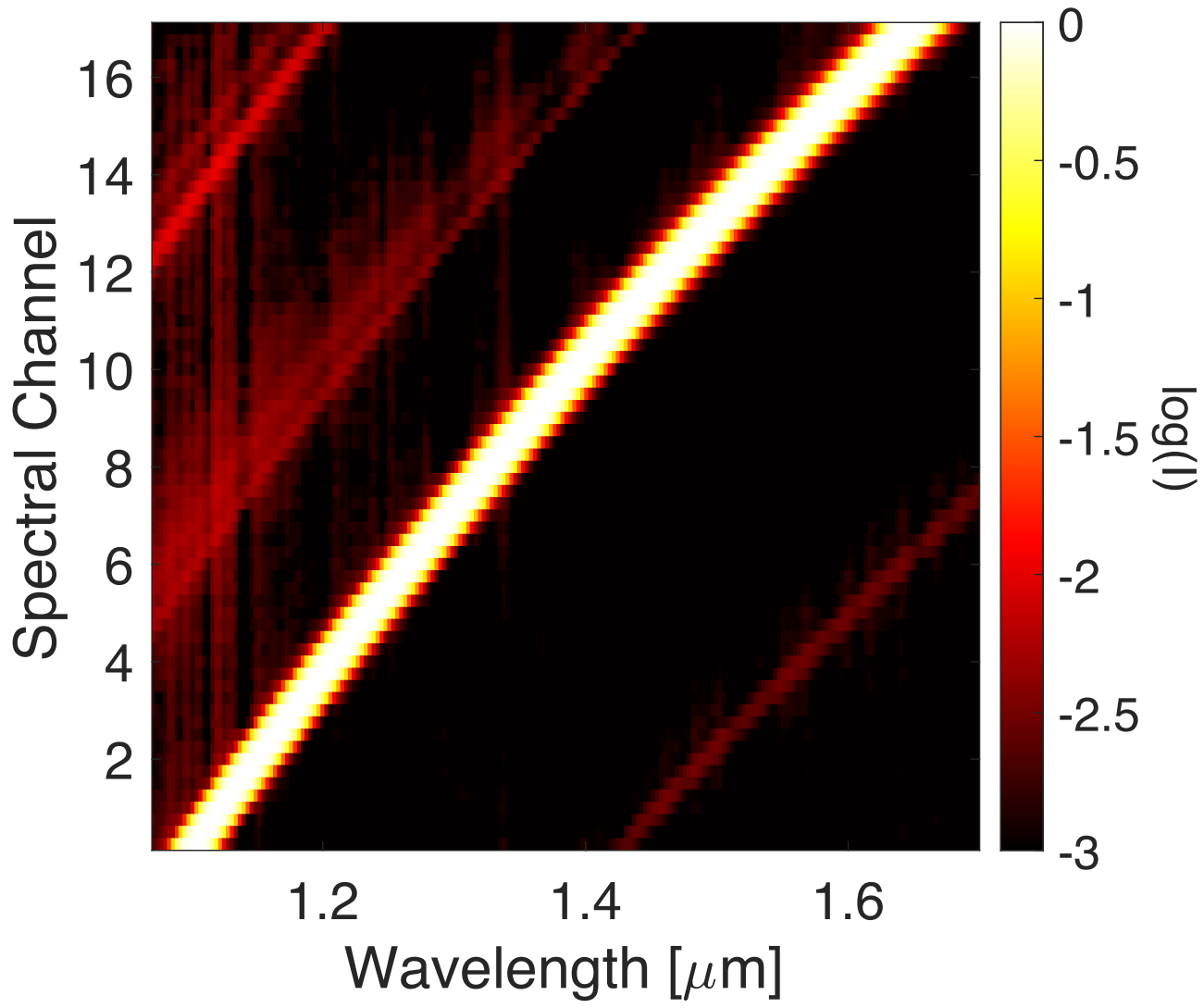

**Figure 20.** Spectral response in Band 2. The horizontal axis shows wavelength and the vertical axis corresponds to detector position, labeled by the nominal spectral channel number used in the survey configuration. The bright diagonal trace marks the band-center progression as the LVF band center varies continuously along the detector. The three off-diagonal features correspond to out-of-band leakage at approximately $0.73\lambda_c$, $0.87\lambda_c$, and $1.3\lambda_c$, with peak amplitudes of order 1%, 0.5%, and 0.3%, respectively.

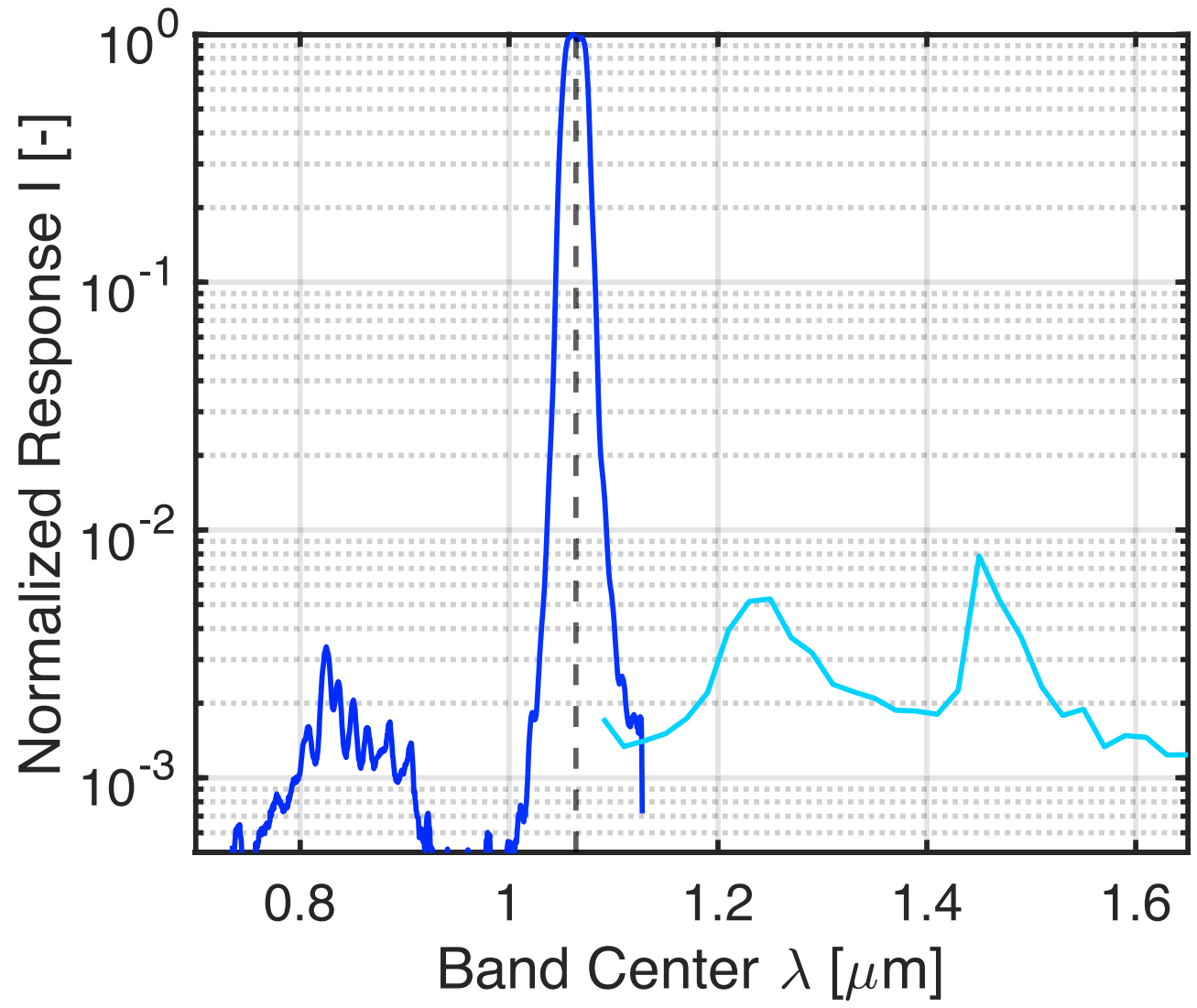

**Figure 21.** Band 1 and 2 detectors respond to 1.064 $\mu$m laser illumination. This figure shows pixels in Band 1 with $\lambda_c = 0.82$ $\mu$m responding to the 1.064 $\mu$m laser at the ~0.3% level, and pixels in Band 2 with $\lambda_c = 1.22$ $\mu$m and 1.46 $\mu$m responding at the ~0.5% and 0.8% level, respectively. These measurements confirm the in-band leakage measured in the spectral calibration campaign using monochromator scans.

Fiducial response functions were computed for the nominal 17 channels per band, as shown in Figure 22, as well as for finer subdivisions into 34, 51, and 68 channels. These products are included in the official SPHEREx data release, allowing users to select the level of spectral binning that best balances resolution and computational efficiency.

### 6.4. In-flight Verification

SPHEREx operates in a Sun-synchronous polar orbit at an altitude of approximately 700 km, where the instrument observes atmospheric emission features in addition to astrophysical backgrounds. A particularly strong feature is the helium airglow line at 1.083 $\mu$m, which is detected in Band 1 and Band 2 and provides a useful on-orbit verification of the ground-based spectral calibration (H. Hui et al. 2026, in preparation). One-dimensional spectra extracted from on-sky observations provide an independent cross-check of the measured band centers near the helium line, as shown in Figure 23.

Additional in-flight verification is obtained from bright astrophysical targets observed in regions with high exposure depth. One example is the Cat's Eye Nebula (J. L. Hora et al. 1999), located near the northern deep fields in the survey, where repeated observations provide dense sampling in wavelength. For this source, we identified all SPHEREx exposures in which the nebula was mapped, totaling of order 1500 exposures per detector band. Rather than binning these measurements into the nominal 102 spectral channels, the data were combined according to the exact wavelength associated with the individual focal plane pixels sampling the source. Figure 24 illustrates the Band 5 focal plane pixels contributing to this analysis and demonstrates the dense wavelength sampling enabled by the repeated coverage. Local background subtraction was performed using nearby sky regions to remove zodiacal-light and diffuse foreground emission. This approach yields an effectively continuous on-sky spectrum that is not limited by the discrete channel definition. The resulting spectrum is shown in Figure 25, enabling a direct comparison between the on-sky measurement and the ground-calibrated instrumental response. The Cat's Eye Nebula is particularly well suited for this analysis because of its location in the northern deep field, where the exposure depth is approximately 400 times higher than that of the all-sky survey.

The measured nebular spectrum shows broad agreement with the ground-based band-center calibration across all bands. At longer wavelengths, small offsets relative to the ground calibration are observed, but these remain within the expected absolute accuracy of the monochromator measurements. Ongoing analysis of additional on-sky data will be used to further refine the wavelength solution, with updates planned for a future data release.

## 7. Conclusion

The spectral calibration campaign for SPHEREx, conducted between 2023 November and 2024 January, produced a complete per-pixel spectral characterization of all six detector bands, spanning the wavelength range 0.75–5.0 $\mu$m. The final data products include the full per-pixel spectral response functions, as well as derived maps of band center and spectral resolution for more than 25 million pixels. Systematic uncertainties associated with the calibration setup and analysis procedure have been quantified. Out-of-band spectral leakage is within design specifications, and nonresponsive pixels identified during this campaign have been incorporated into the permanent bad-pixel mask for mission operations.

The integrated telescope, including detectors, LVFs, and dichroic beam splitters, satisfies the spectral performance targets established in the mission design. The fiducial spectral response functions provide an efficient representation of the instrument response for routine science analysis, while the full per-pixel dataset supports applications that require the highest available spectral accuracy. The calibration products described here are part of the SPHEREx data release, enabling reproducible analysis and future reprocessing of the full mission dataset.

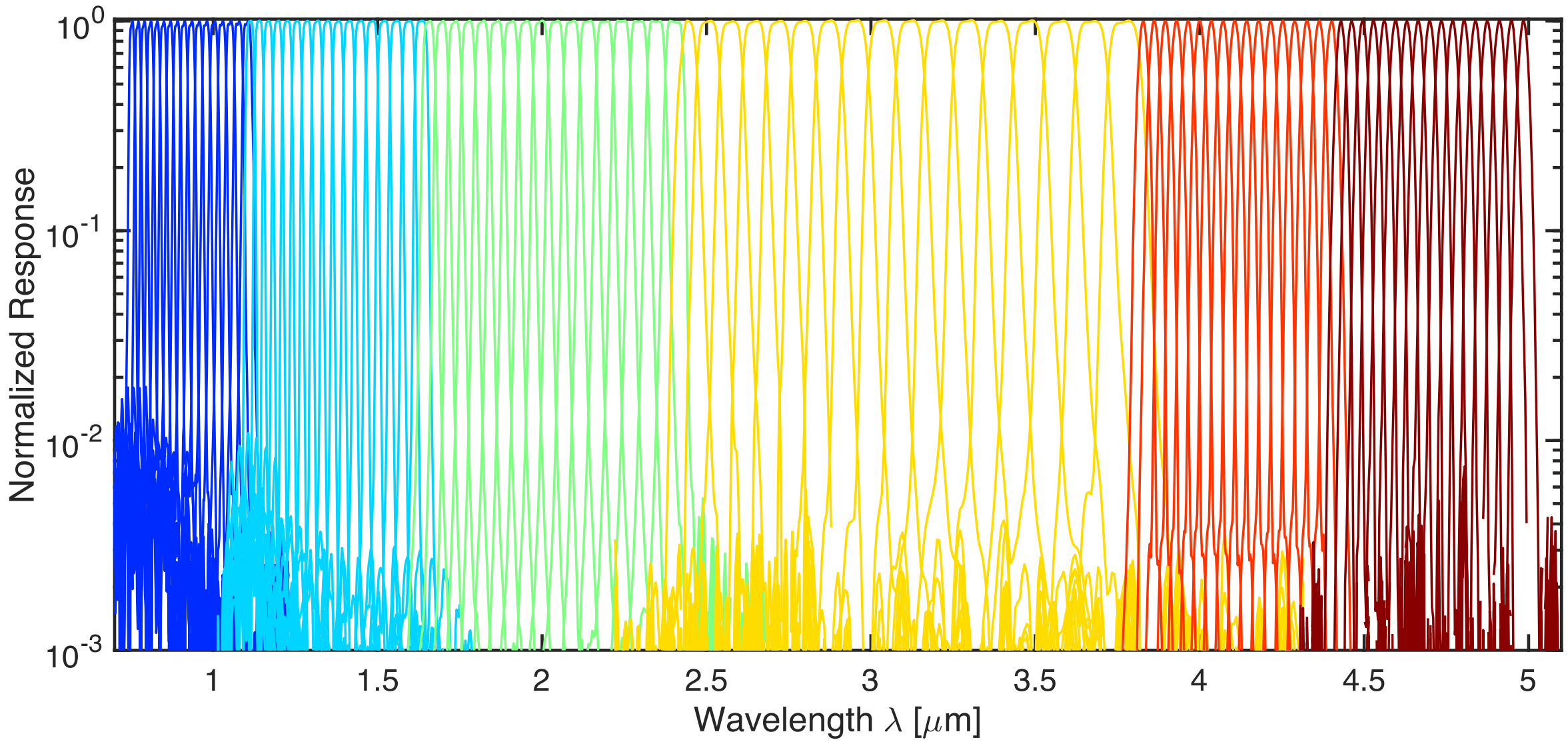

**Figure 22.** Fiducial spectral response functions for all six detectors with nominal 17 channels per detector, showing overlapping spectral coverage between bands and minimal out-of-band leakage.

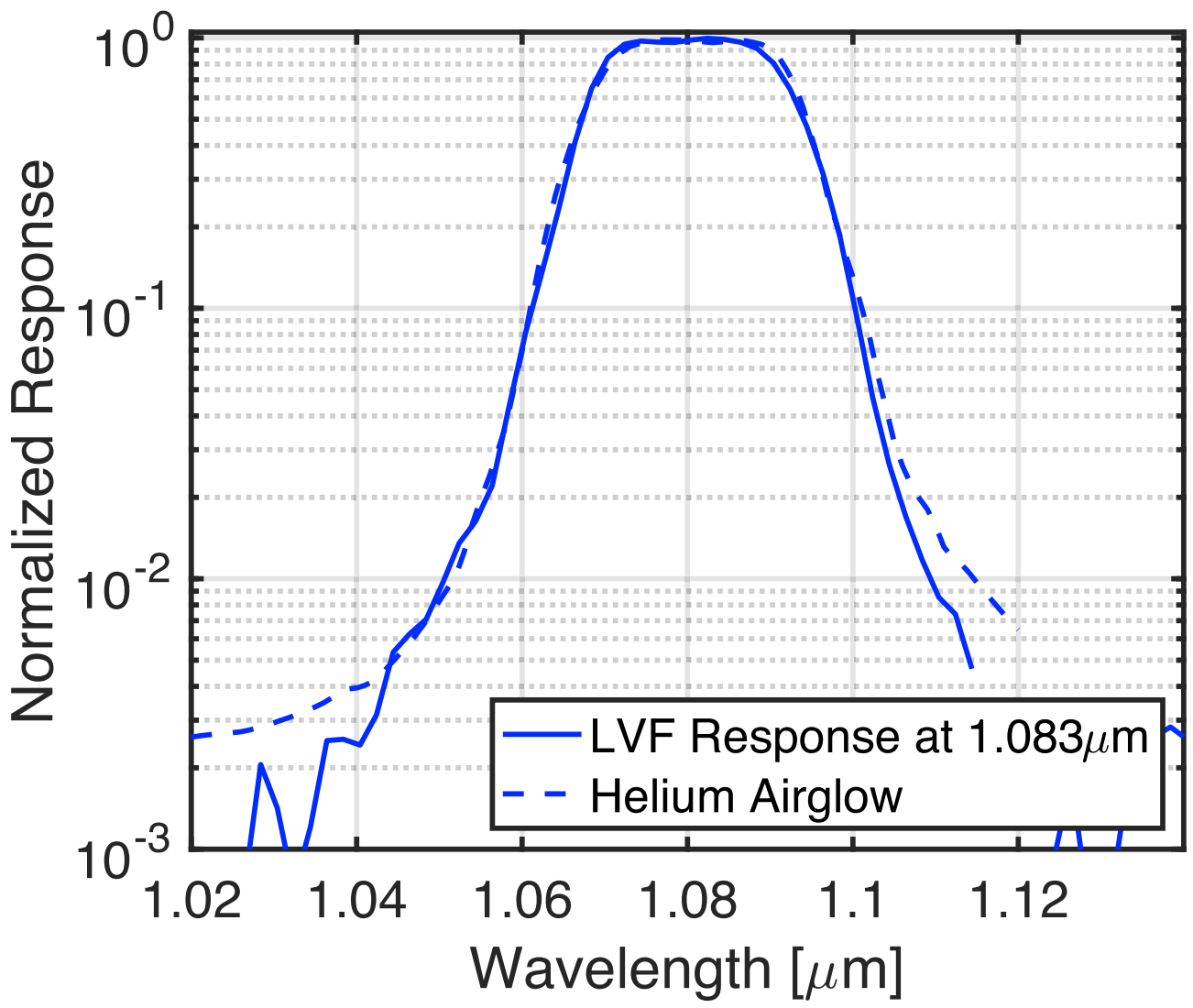

**Figure 23.** Helium airglow emission at 1.083 $\mu$m observed in Band 1, overlaid with the expected instrument response for a monochromatic 1.083 $\mu$m emission line based on the ground calibration. The measured and predicted line positions agree to better than 0.1 nm. The slightly broader airglow feature in the flight data is due to imperfect zodiacal-light subtraction.

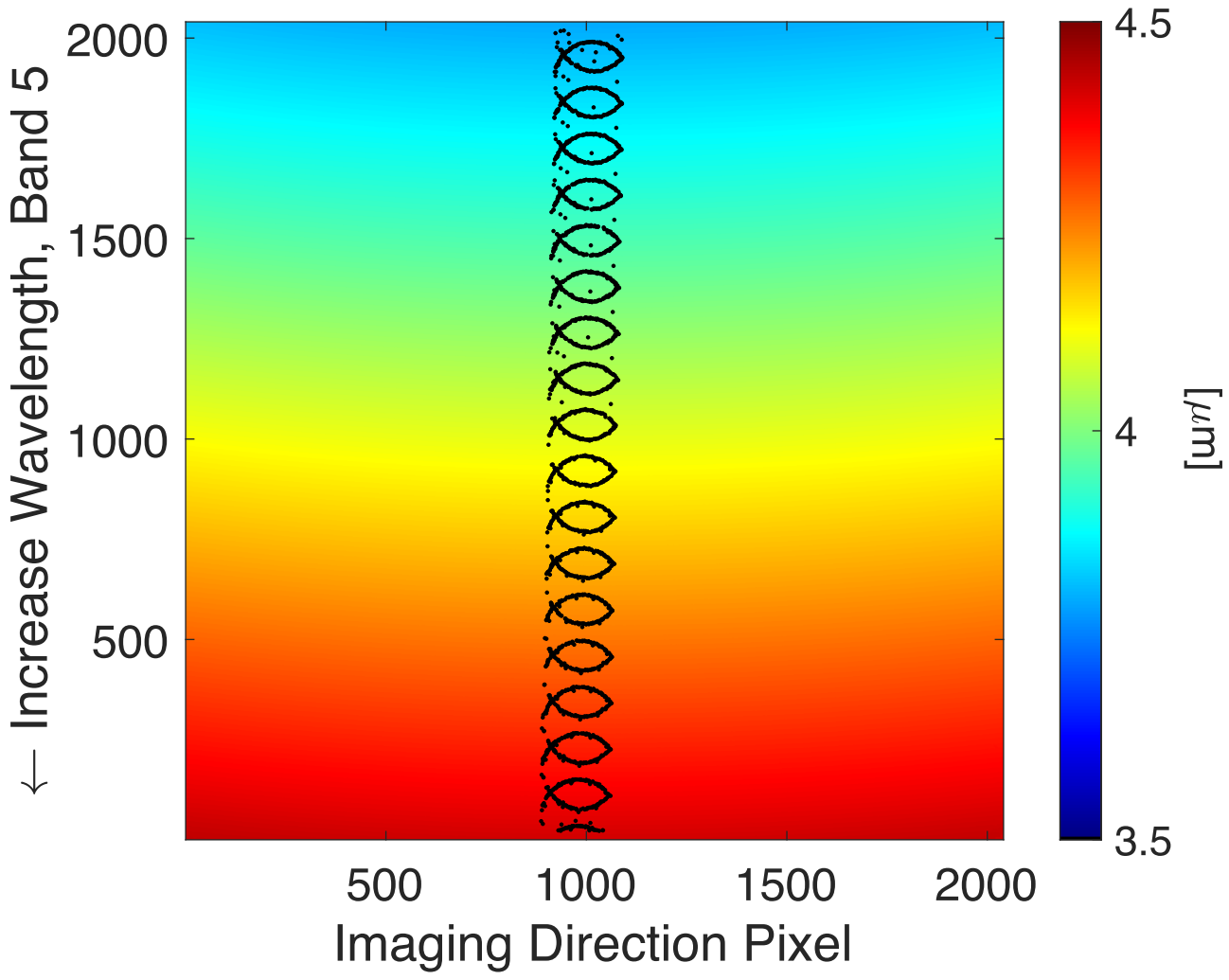

**Figure 24.** Band 5 focal plane sampling of the Cat's Eye Nebula. The background image shows the detector wavelength map, with color indicating the center wavelength of each pixel from 3.82 to 4.42 $\mu$m. Black points correspond to individual SPHEREx exposures, with each point marking the focal plane pixel in which the Cat's Eye Nebula is mapped for that exposure. Because the wavelength varies with pixel position, the ensemble of measurements provides dense sampling of the nebular spectrum. Equivalent mappings are performed for all bands, and together these measurements produce the on-sky spectrum shown in Figure 25.

Early in-flight measurements, including atmospheric helium emission and deep field observations of bright nebular targets, confirm the ground-based wavelength solution to within 1 nm in Bands 1 through 3, and to approximately 10 nm in Bands 4 to 6. Additional refinements based on accumulated in-flight observations will be incorporated into updated spectral response models as needed, particularly for the long-wavelength bands.

## Acknowledgments

We acknowledge support from the SPHEREx project under a contract from the NASA/Goddard Space Flight Center to the California Institute of Technology.

Part of the research described in this paper was carried out at the Jet Propulsion Laboratory, California Institute of Technology, under a contract with the National Aeronautics and Space Administration (80NM0018D0004).

The authors acknowledge the Texas Advanced Computing Center (TACC) at The University of Texas at Austin for providing computational resources that have contributed to the research results reported within this paper.

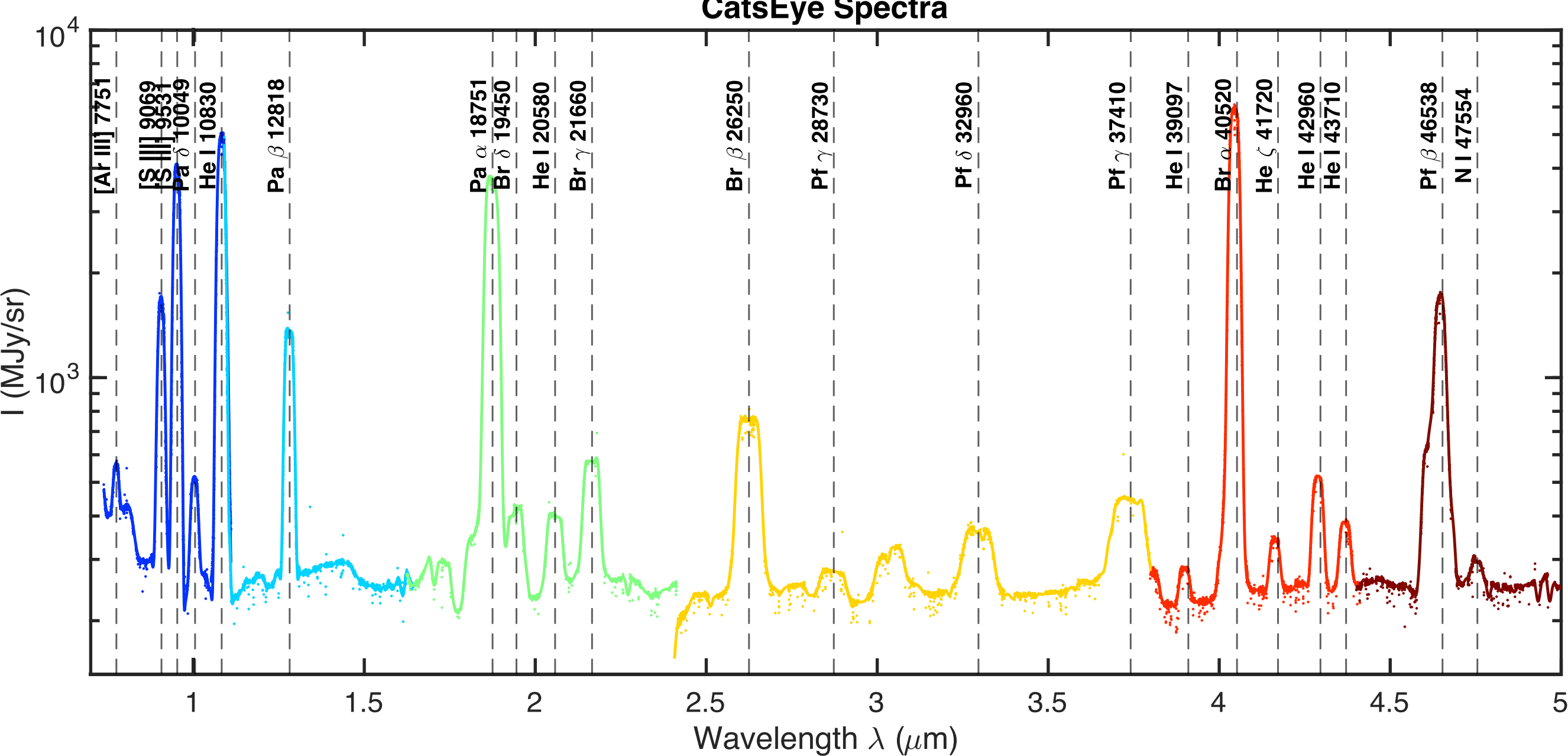

**Figure 25.** Spectrum of the Cat's Eye Nebula observed in a deep survey field. The high redundancy in wavelength sampling makes this source useful for in-flight verification of the spectral response model.

## Appendix
## Calibration Hardware and Optical Design

This appendix provides additional details of the warm optical bench, cryogenic chambers, and calibration optics used in the SPHEREx spectral calibration campaign.

### *A.1. Warm Optical Bench*

#### *A.1.1. Calibration Light Sources*

Multiple illumination sources were used to cover the full SPHEREx spectral range. A Thorlabs SLS301, intensity stabilization tungsten halogen lamp[12] served as the primary broadband source for all six detector bands. At longer wavelengths, a Thorlabs SLS203F silicon carbide Globar lamp[13] provided higher output and was used as an independent cross-check to verify that the measured bandpasses in Bands 5 and 6 were consistent with those obtained using the SLS301 tungsten halogen source.

For absolute wavelength calibration and systematic verification, several atomic emission line sources were employed, including a Newport xenon arc lamp[14] and pencil-style krypton, neon, and argon calibration lamps.[15] These lamps produce narrow, known spectral lines that were used to validate the wavelength solution and assess systematic uncertainties, as described in Section 5.1.

Two fixed-wavelength lasers at 1.064 $\mu$m and 1.550 $\mu$m were also used to validate the in-band spectral response in Bands 1 and 2.

[12] https://www.thorlabs.com/thorproduct.cfm?partnumber=SLS301
[13] https://www.thorlabs.com/thorproduct.cfm?partnumber=SLS203F
[14] https://www.newport.com/f/high-power-xenon-research-light-sources
[15] https://www.newport.com/f/pencil-style-calibration-lamps

#### *A.1.2. Monochromator*

A Newport Oriel CS260 monochromator[16] with $f/3.9$ optics was used to generate tunable monochromatic light for the calibration. The monochromator contains an entrance slit, a motorized turret that houses three interchangeable diffraction gratings, and an exit slit. A pair of $CaF_2$ lenses was installed at the output to produce a 2 inch diameter collimated beam. This beam was routed through a set of gold-coated relay mirrors and focused by a parabolic mirror onto a pinhole located inside the cryogenic chamber.

The spectral bandwidth of the monochromator output is set by the grating dispersion and by the selected entrance and exit slit widths. Table 4 summarizes the grating configurations used during the SPHEREx calibration for different portions of the wavelength range.

For Bands 1 through 4, the monochromator bandwidth is roughly an order of magnitude narrower than the intrinsic instrument bandpass, and the illumination can be treated as effectively monochromatic. For Bands 5 and 6, narrow slits do not provide enough throughput because the lamp brightness becomes much weaker relative to the 300 K thermal background at long wavelengths. To preserve the signal-to-noise ratio, wider slits were used, which increased the monochromator output bandwidth to within a factor of 2 of the SPHEREx bandpass. This broadening was modeled and deconvolved in the analysis, as described in Section 4.3.

#### *A.1.3. Order-sorting Filters*

Higher-order diffraction from the monochromator gratings was suppressed using long-pass order-sorting filters installed between the broadband light source and the monochromator entrance slit. The filters were mounted in a motorized

[16] https://www.newport.com/f/cs260-uv-vis-monochromators

**Table 4**
Monochromator Grating and Slit Configurations Used in the Calibration

| Grating No. | 1 | 2 | 3 |
|---|---|---|---|
| Wavelength Range ($\mu$m) | 0.48–1.40 | 0.93–2.60 | 2.50–12.00 |
| Part No. | 74064 | 74071 | 74080 |
| Dispersion (nm mm$^{-1}$) | 3.1 | 9.6 | 25.8 |
| Slits ($\mu$m) | 760 | 500 | 760[a] |

**Notes.** Gratings 1 and 2 provide high-resolution illumination for Bands 1–4. Grating 3 is used for Bands 5 and 6 where a wider slit increases throughput at the longest wavelengths.
[a] Wider slits are used with grating 3 to increase throughput for Bands 5 and 6.

**Table 5**
Order-sorting Filter Configuration

| Filter Slot | 1 | 2 | 3 | 4 | 5 | 6 |
|---|---|---|---|---|---|---|
| Cutoff ($\mu$m) | 0.60 | 1.30 | 1.65 | 2.40 | 3.60 | ... |
| Part No. | 62985 | 89664 | 68652 | 68653 | 68654 | ... |

**Note.** Slot 6 is intentionally left empty.

Newport USFW-100 filter wheel.[17] Filters were chosen to block shorter-wavelength light that would otherwise propagate through higher diffraction orders, while transmitting the desired spectral band.[18]

Each monochromator grating configuration is paired with a specific order-sorting filter to ensure that only the intended diffraction order reaches the monochromator exit slit. Table 5 lists the filters mounted in each wheel slot along with their nominal cutoff wavelengths. Slot 6 is intentionally left empty for configurations where no order-sorting filter is required. The filter selections used for the calibration measurements are summarized in Table 2.

#### *A.1.4. Warm Optics Enclosure and Nitrogen Gas Purge*

The warm optical bench was enclosed within a sealed housing constructed from AmStat antistatic shielding material.[19] This housing was purged with $N_2$ to suppress atmospheric $CO_2$ and water absorption features that would otherwise introduce systematic errors in the spectral calibration.

Prior to each calibration run, the enclosure was continuously purged with dry nitrogen gas for a minimum of 2 hr. A commercial $CO_2$ sensor placed inside the enclosure verified the purge, with $CO_2$ concentrations reduced to less than 5 ppm, compared to ambient room levels of approximately 400 ppm.

### *A.2. Cryogenic Chambers and Calibration Optics*

#### *A.2.1. Cryogenic Test Chambers*

A small cryostat was used for testing the stand-alone focal plane assemblies and other subsystems prior to telescope integration. The cryostat has a Cryomech PT-815[20] pulse tube cryo-cooler capable of reaching a 20 K base temperature. Heaters mounted on the 20 K cold plate are controlled using two LakeShore Model 336 cryogenic temperature controllers.[21] Closed-loop temperature regulation is implemented with a modified version of PyHK, a Python-based control package developed for cryogenic test beds (J. Hunacek 2020), enabling stable operation between 38 K and 120 K with better than 1 mK stability during spectral scans.

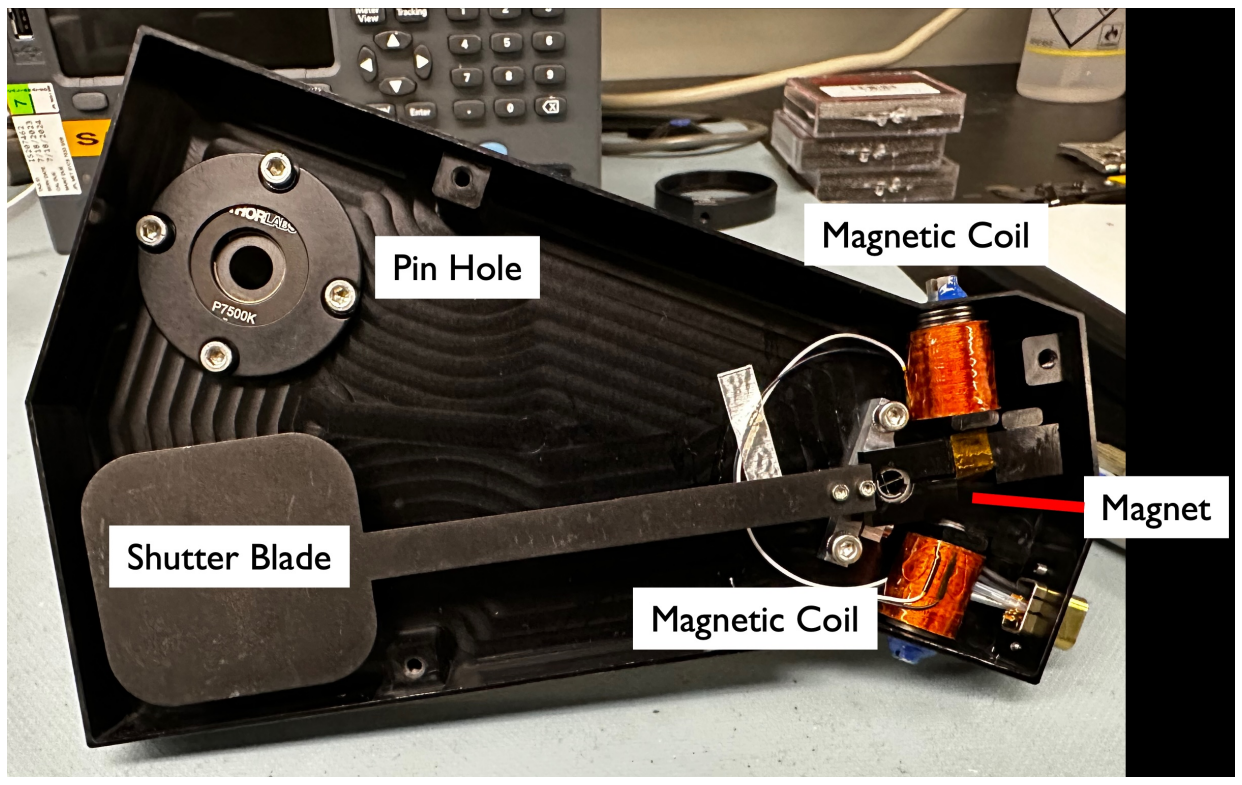

**Figure 26.** Cryogenic shutter assembly. The blade is actuated by two magnetic coils and is machined from anodized black aluminum with an Acktar coating on the rear surface to suppress stray light.

The focal plane assembly was mounted directly onto the 20 K cold plate, and the Winston cone and integrating spheres were installed on the same stage inside an optically blackened enclosure surrounding the detector. This enclosure suppressed stray light to ensure that only the monochromatic beam reached the focal plane. A 2 inch sapphire optical window at the bottom of the chamber coupled the monochromatic beam from the warm optical bench into the cold volume.

System-level calibration was carried out in a larger cryogenic chamber developed by KASI. This chamber accommodated the fully assembled SPHEREx telescope and maintained the required 45–62 K thermal environment for end-to-end spectral characterization (W.-S. Jeong 2024). The chamber is cooled by two Cryomech PT-815 units. The rear section of the chamber housed the telescope, while the front section contained the cryogenic shutter, integrating spheres, and Winston cone used to illuminate the telescope entrance pupil. A sapphire optical window identical to that used in the focal plane test bed coupled the monochromatic beam into the chamber.

#### *A.2.2. Cryogenic Shutter*

The cryogenic shutter assembly is mounted immediately in front of the entrance to the first integrating sphere. Inside the shutter housing is a magnetically actuated blade. The shutter provides a method for acquiring dark exposures during calibration. The mechanical design was adapted from the shutter developed for the CIBER-2 experiment (I. S. Sullivan 2011; M. Zemcov et al. 2025). A photograph of the assembled unit is shown in Figure 26.

The shutter blade is fabricated from thin anodized black aluminum and coated with Acktar tape to suppress reflections and stray light. Two electromagnetic coils are mounted on opposite sides of the assembly actuate the blade. A brief current pulse drives the blade to snap between the open and closed positions, and the blade remains mechanically latched without any holding current. This eliminates steady-state power dissipation at cryogenic temperature and avoids thermal disturbances during data acquisition.

[17] https://www.newport.com/p/USFW-100
[18] https://www.edmundoptics.com/f/infrared-ir-longpass-filters/14164/
[19] https://www.uline.com/BL_54/Static-Shielding-Rolls
[20] https://bluefors.com/products/pulse-tube-cryocoolers/pt815-pulse-tube-cryocooler/
[21] https://www.lakeshore.com/products/categories/overview/temperature-products/cryogenic-temperature-controllers/model-336-cryogenic-temperature-controller

**Table 6**
Physical Dimensions of the Integrating Spheres and Winston Cones Used in the Focal Plane and Telescope Configurations

| Component | FPA Config. | Telescope Config. |
|---|---|---|
| *First Integrating Sphere* | | |
| Sphere Radius (mm) | 50 | 100 |
| Input Port Radius (mm) | 2 | 5 |
| Output Port Radius (mm) | 2 | 5 |
| *Second Integrating Sphere* | | |
| Sphere Radius (mm) | 50 | 400 |
| Input Port Radius (mm) | 2 | 5 |
| Output Port Radius (mm) | 2 | 127 |
| *Winston Cone* | | |
| Length (mm) | 489 | 915 |
| Input Port Radius (mm) | 2 | 127 |
| Output Port Radius (mm) | 2 | 300 |
| Designed $f/\#$ (mm) | 3 | 1.07 |

Electrical power is delivered through hermetically sealed vacuum feedthroughs from an external supply located outside the chamber. The same shutter assembly was used in both the focal plane test configuration and the full telescope calibration.

#### A.2.3. Integrating Spheres and Winston Cone

After passing through the shutter assembly, the monochromatic beam is focused onto the pinhole located at the shutter entrance. Downstream of the pinhole, two integrating spheres homogenize the spatial intensity distribution of the beam. Two spheres are required because the output ports of a single integrating sphere are fixed at 90° relative to the input port, whereas the calibration optics inside the cryogenic chamber must maintain a straight optical path aligned with the chamber geometry for mechanical simplicity.

The focal plane test configuration has two Newport 819D-GL-3[22] integrating spheres, while the system-level calibration employs large custom spheres built by KASI. Key geometric parameters for both configurations are summarized in Table 6.

The output of the second integrating sphere couples directly into a cryogenic Winston cone (R. Winston 1970). The Winston cone functions as a nonimaging concentrator: it preserves etendue while reshaping the diffuse integrating sphere output into a well-defined beam matched to the instrument focal ratio. The cone geometry is defined by its length and by the radii of its input and output apertures, with the surface profile described by

$$\begin{aligned} 0 = {} & r^2 \cos^2\theta \\ & + r[2z\sin\theta\cos\theta + 2a(1+\sin\theta)^2] \\ & + z^2\sin^2\theta - 2az\cos\theta(2+\sin\theta) \\ & - a^2(3+\sin\theta)(1+\sin\theta), \end{aligned} \tag{A1}$$

where $r$ is the cone radius at axial position $z$, $a$ is the input aperture radius, and $\theta$ is the acceptance angle.

The Winston cone is designed to produce a spatially uniform beam that covers the full SPHEREx field of view. In practice, we observe small asymmetries introduced by internal baffles between the ports of the integrating spheres, as shown in Figure 7. These spatial variations are not critical for the spectral calibration, as their structure does not change much with wavelength.

Thermal control of the assembly is maintained by mounting the integrating spheres, internal baffles, and Winston cone onto the same temperature stage as the instrument. This ensures that all elements remain at a common temperature throughout the calibration sequence. During the telescope campaign in the KASI chamber, additional thermal straps were added between the optical components and the chamber cold plate to accelerate cooldown and improve temperature uniformity. The full integrating sphere and Winston cone assembly reached approximately 80 K during measurements, sufficiently cold to keep the thermal background below the detector dark current level.

Following Labsphere (2017), the efficiency of an integrating sphere is

$$\eta_{\rm sphere} = \frac{1}{A_{\rm sphere}} \frac{\rho}{1-\rho(1-f)} \pi R_{\rm out}^2, \tag{A2}$$

where $R_{\rm out}$ is the output port radius, $A_{\rm sphere}$ is the surface area of the sphere, $\rho$ is the wall reflectance, and the port fraction $f$ is

$$f = \frac{A_{\rm in} + A_{\rm out}}{A_{\rm sphere}}, \tag{A3}$$

with $A_{\rm in}$ and $A_{\rm out}$ the areas of the input and output ports, respectively.

Assuming negligible coupling loss between the second integrating sphere and the Winston cone, the power received by a single pixel in the focal plane configuration is

$$P_{\rm FPA} \simeq P_{\rm in} \cdot \frac{\eta_{\rm sphere1}\,\eta_{\rm sphere2}}{A_{\rm winston}} \cdot \eta_{\rm det} \cdot A_{\rm pix}, \tag{A4}$$

where $\eta_{\rm sphere1}$ and $\eta_{\rm sphere2}$ are the efficiencies of the two spheres, $\eta_{\rm det}$ is the LVF plus detector efficiency, $P_{\rm in}$ is the input optical power, and $A_{\rm pix}$ and $A_{\rm winston}$ are the pixel area and Winston cone output area, respectively.

Similarly, the power received by a single pixel in the telescope configuration is

$$P_{\rm tel} \simeq P_{\rm in} \cdot \frac{\eta_{\rm sphere1}\,\eta_{\rm sphere2}}{A_{\rm winston}\,\Omega_{\rm winston}} \cdot \eta_{\rm det} \cdot A_{\rm tel} \cdot \Omega_{\rm pix}, \tag{A5}$$

where $A_{\rm tel}$ is the telescope aperture, and $\Omega_{\rm pix}$ and $\Omega_{\rm winston}$ are the pixel and Winston cone output solid angle, respectively.

Although the absolute power output of the monochromator is difficult to predict due to uncertainties in grating efficiency and coupling into the integrating sphere, comparisons with the measured photocurrent under 300 K background illumination show agreement with the model, as summarized in Figure 8. This agreement underpins the thermal background discussion in Section 3.3.

## ORCID iDs

Howard Hui https://orcid.org/0000-0001-5812-1903
James J. Bock https://orcid.org/0000-0002-5710-5212
Samuel Condon https://orcid.org/0000-0003-4255-3650
C. Darren Dowell https://orcid.org/0009-0002-0098-6183
Woong-Seob Jeong https://orcid.org/0000-0002-2770-808X
Young-soo Jo https://orcid.org/0000-0003-3574-1784

[22] https://www.newport.com/f/819d-gold-integrating-spheres

Phil M. Korngut https://orcid.org/0009-0003-8869-3365
Kenneth Manatt https://orcid.org/0009-0009-5777-4457
Chi Nguyen https://orcid.org/0000-0001-9368-3186
Stephen Padin https://orcid.org/0009-0001-9993-4393
Sung-Joon Park https://orcid.org/0009-0001-4064-9918
Jeonghyun Pyo https://orcid.org/0000-0001-9937-8270
Yujin Yang https://orcid.org/0000-0003-3078-2763
Matthew L. N. Ashby https://orcid.org/0000-0002-3993-0745
Yoonsoo P. Bach https://orcid.org/0000-0002-2618-1124
Tzu-Ching Chang https://orcid.org/0000-0001-5929-4187
Yun-Ting Cheng https://orcid.org/0000-0002-5437-0504
Yi-Kuan Chiang https://orcid.org/0000-0001-6320-261X
Asantha Cooray https://orcid.org/0000-0002-3892-0190
Brendan P. Crill https://orcid.org/0000-0002-4650-8518
Ari J. Cukierman https://orcid.org/0000-0002-7471-719X
Olivier Doré https://orcid.org/0000-0001-7432-2932
Andreas L. Faisst https://orcid.org/0000-0002-9382-9832
Joseph L. Hora https://orcid.org/0000-0002-5599-4650
Jae Hwan Kang https://orcid.org/0000-0002-3470-2954
BoMee Lee https://orcid.org/0000-0003-1954-5046
Carey M. Lisse https://orcid.org/0000-0002-9548-1526
Daniel C. Masters https://orcid.org/0000-0001-5382-6138
Roberta Paladini https://orcid.org/0000-0002-5158-243X
Zafar Rustamkulov https://orcid.org/0000-0003-4408-0463
Volker Tolls https://orcid.org/0000-0003-1841-2241
Michael W. Werner https://orcid.org/0000-0003-4990-189X
Michael Zemcov https://orcid.org/0000-0001-8253-1451